\documentclass[%
prx,
twocolumn,
english,
superscriptaddress,
floatfix,
amsmath,amssymb,
longbibliography
]{revtex4-2}
\usepackage{tikz}
\usetikzlibrary{circuits.ee.IEC}
\usepackage{graphicx}
\usepackage{dcolumn}
\usepackage{soul}
\usepackage{bm}
\usepackage[hidelinks]{hyperref}
\usepackage{siunitx}
\hypersetup{colorlinks = true, allcolors = blue}

\usepackage{amssymb} 
\usepackage{amsmath} 
\usepackage{bm}
\usepackage{mathtools}
\usepackage{xcolor}

\newcommand{\jbfern}[1]{\textcolor{magenta}{#1}}

\newcommand{\te}[1]{\mbox{$\mathbf{ #1 }$}}

\def\eq{\ = \ }

\def\bnabla{\boldsymbol{\nabla}}

\def\bbe{\te{e}}

\def\bj{\te{j}}

\def\bn{\te{n}}

\def\bx{\te{x}}

\def\bD{\te{D}}

\def\b1{\te{1}}

\graphicspath{{}}

\begin{document}

\righthyphenmin=4
\lefthyphenmin=4


\title{Spatiotemporal dynamics of ionic reorganization \\ near biological membrane interfaces}

\author{Hyeongjoo Row}
\thanks{Equal contribution}
\affiliation{Department of Chemical and Biomolecular Engineering, University of California, Berkeley, CA 94720, USA\looseness=-1}
\affiliation{Helen Wills Neuroscience Institute, California Institute for Quantitative Biosciences, QB3, Center for Computational Biology, University of California, Berkeley, CA 94720, USA}

\author{Joshua B. Fernandes}
\thanks{Equal contribution}
\affiliation{Department of Chemical and Biomolecular Engineering, University of California, Berkeley, CA 94720, USA\looseness=-1}

\author{Kranthi K. Mandadapu}
\thanks{Correspondence: kranthi@berkeley.edu, kshekhar@berkeley.edu}
\affiliation{Department of Chemical and Biomolecular Engineering, University of California, Berkeley, CA 94720, USA\looseness=-1}
\affiliation{Chemical Sciences Division, Lawrence Berkeley National Laboratory, CA 94720, USA}

\author{Karthik Shekhar}
\thanks{Correspondence: kranthi@berkeley.edu, kshekhar@berkeley.edu}
\affiliation{Department of Chemical and Biomolecular Engineering, University of California, Berkeley, CA 94720, USA\looseness=-1}
\affiliation{Helen Wills Neuroscience Institute, California Institute for Quantitative Biosciences, QB3, Center for Computational Biology, University of California, Berkeley, CA 94720, USA}
\affiliation{Biological Systems Division, Lawrence Berkeley National Laboratory, Berkeley, CA 94720, USA}

\date{\today}

\begin{abstract}
Electrical signals in excitable cells involve spatially localized ionic fluxes through ion channels and pumps on cellular lipid membranes. 
Common approaches to understand how these localized fluxes spread assume that the membrane and the surrounding electrolyte comprise an equivalent circuit of capacitors and resistors, which ignores the localized nature of transmembrane ion transport, the resulting ionic gradients and electric fields, and their spatiotemporal relaxation. 
Here, we consider a model of localized ion pumping across a lipid membrane, and use theory and simulation to investigate how the electrochemical signal propagates spatiotemporally in- and out-of-plane along the membrane.
The localized pumping generates long-ranged electric fields with three distinct scaling regimes along the membrane: a constant potential near-field region, an intermediate ``monopolar" region, and a far-field ``dipolar" region. Upon sustained pumping, the monopolar region expands radially in-plane with a steady speed that is enhanced by the dielectric mismatch and the finite thickness of the lipid membrane. 
For unmyelinated lipid membranes in physiological settings, we find remarkably fast propagation speeds of $\sim\!40 \, \mathrm{m/s}$, allowing faster ionic reorganization compared to bare diffusion.
Together, our work shows that transmembrane ionic fluxes induce transient long-ranged electric fields in electrolyte solutions, which may play hitherto unappreciated roles in biological signaling.

\end{abstract}


\maketitle

\section{Introduction}

\begin{figure}[t]
\centering
\includegraphics[width=0.95\linewidth]{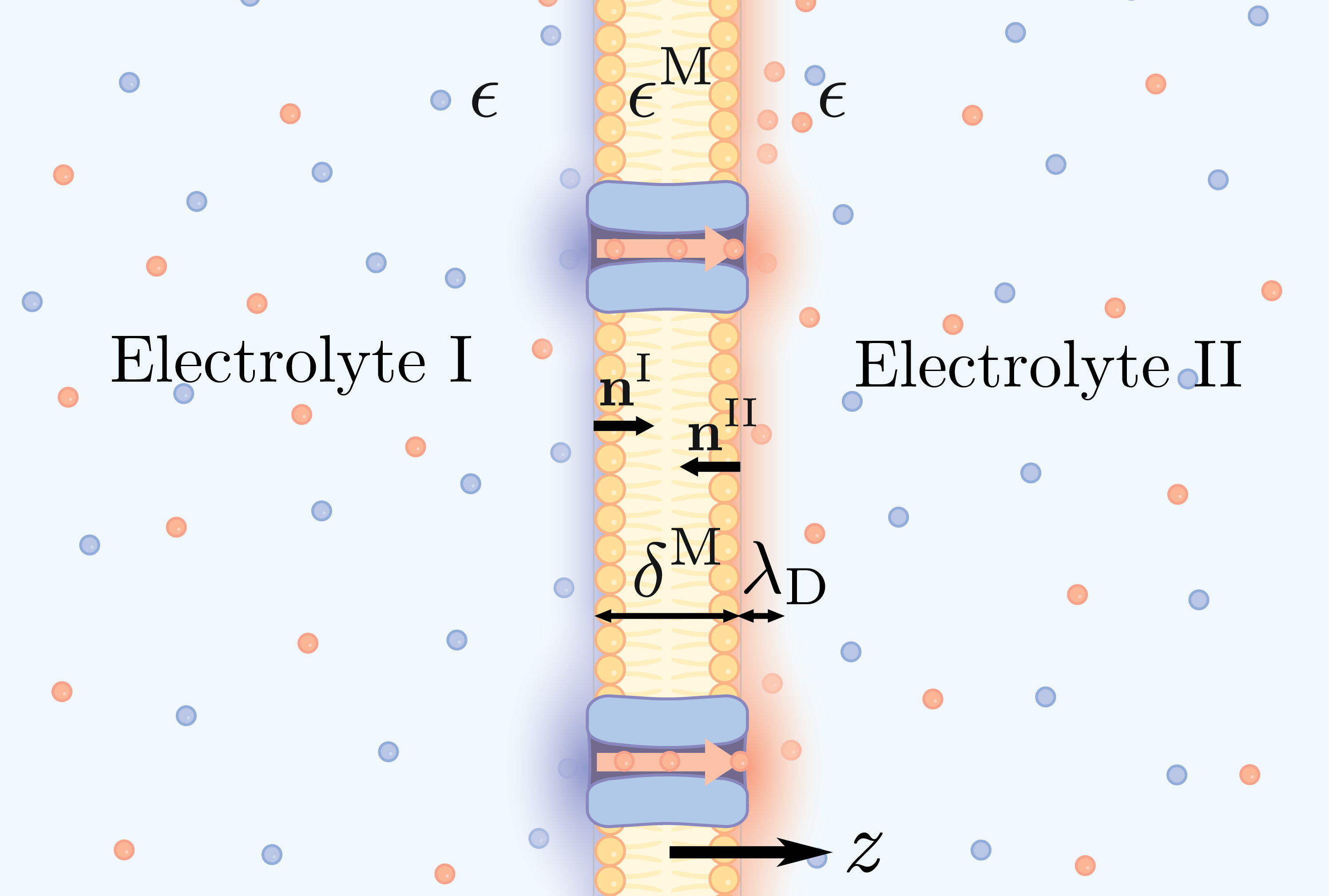}    
\caption{A lipid membrane of thickness $\delta^{\rm M}$ and dielectric permittivity $\epsilon^{\rm M}$ separates two dilute aqueous electrolytes I (${z<-\delta^{\rm M}/2}$) and II (${z>\delta^{\rm M}/2}$) with dielectric permittivity $\epsilon$. Red and blue circles indicate cations and anions, respectively. Transmembrane pumps transport cations from the domain $\rm I$ to $\rm II$. The selective pumping breaks electroneutrality, leading to interfacial diffuse charge layers of thickness $\lambda_{\rm D}$, the Debye length. Outward normals of domains $\rm I$ and $\rm II$ at the membrane interfaces are denoted $\bn^{\rm I}$ and $\bn^{\rm II}$, respectively.}
\label{fig1_schematic_descriptions}
\end{figure}

Electrical signaling in excitable cells such as neurons and heart muscle cells relies on the selective partitioning and transport of ions across lipid membranes~\cite{lodish2000molecular,hille1992,hodgkin1949effect,hodgkin1952quantitative}, resulting in the dynamic regulation of $\Delta V^{\rm M}$, the transmembrane electric potential. This potential $\Delta V^{\rm M}$ is what controls the voltage-dependent activity of membrane proteins such as ion channels and pumps, which in turn regulate the selective movement of ions \cite{hille1992}. Ionic fluxes through these proteins create localized concentration perturbations, which are followed by relaxation processes involving diffuse charge reorganization at the membrane-fluid interface (Fig.~\ref{fig1_schematic_descriptions}).

The classical approach to model electrical signaling, \emph{cable theory}, represents a neuron as an equivalent electrical circuit \cite{cole1938,hodgkin1946,rall1962}. Cable models treat nerve fibers as conductors, the extracellular and intracellular spaces as perfect resistors, and the cell membrane as a capacitor that permits some conductance \cite{dayan2001}. These models assume microscopic ionic homogeneity, i.e. electroneutrality, and neglect spatiotemporal ionic dynamics by considering only the evolution of the transmembrane potential.  However, it is known that electroneutrality is violated during active currents through membranes and at the nanometer length scales, either close to membrane interfaces or in restricted biological spaces such as synaptic clefts and dendritic spines \cite{savtchenko2017electrodiffusion,sylantyev2008,cartailler2018deconvolution}. This breakdown of electroneutrality produces variations in local electric fields, which in turn drive ionic reorganization. Furthermore, the presence of the lipid membrane, which has a finite thickness ($3\,\text{-}\,7\, \text{nm}$) \cite{lodish2000molecular} and a ${\sim20}$-fold lower dielectric permittivity compared to the bulk aqueous solution \cite{huang1977theoretical,nymeyer2008method}, is commonly represented by a single capacitance. How the spatiotemporal propagation of ionic fluxes and the large dielectric mismatch affect in-plane ionic reorganization, and their downstream consequences for signal propagation, remain poorly understood.

The equivalent circuit and electroneutral assumptions of cable theory may be addressed using the Poisson-Nernst-Planck (PNP) framework \cite{Nernst1888,Nernst1889,Planck1890,Bazant04}, i.e. the dilute solution limit of transport phenomena in electrolyte solutions \cite{fong2020transport}, which has been extensively used to model charge dynamics in electrochemical systems~\cite{tedesco2016nernst,mei2018physical,wu2022understanding}. In the context of biological systems, PNP equations have been utilized to model ion transport through individual ion channels \cite{chen1997,liu2014,singer2008}. More recently, there has been a resurgence of interest in the PNP framework among neurophysiologists, who, acknowledging the limitations of cable models, used PNP equations to study the clearance of charged neurotransmitters in synaptic clefts \cite{savtchenko2004electric}, synaptic currents in dendritic spines \cite{cartailler2017,lagache2019}, and the propagation of action potentials under the assumption of Hodgkin-Huxley-type membrane conductances \cite{pods2013electrodiffusion, gulati2023}. Nevertheless, these existing works primarily consider either 1D models that neglect in-plane phenomena at the membrane interface or are limited to numerical studies in the 2D and 3D settings still reliant on equivalent circuit ideas. This leaves open questions about the roles of the breakdown of electroneutrality, the membrane thickness, and the dielectric mismatch in the spatiotemporal reorganization of ions even within the PNP framework.

In this work, we study the spatiotemporal aspects of ionic reorganization by considering the dynamics induced by a single ion transporter embedded in a lipid membrane characterized by a finite thickness and dielectric mismatch with the electrolyte solutions (Fig.~\ref{fig1_schematic_descriptions}). This serves as a classical source-sink problem with the transporter creating a source and a sink of charge on each side of the membrane. By combining large-scale numerical simulations and analytical solutions of the PNP equations, we show that charge separation resulting from selective pumping of a single ionic species leads to long-range electric fields. These fields drive ionic reorganization between the bulk electrolyte solution and a diffuse charge layer (also known as the electrical double layer) along the membrane surface where electroneutrality is violated. This mode of reorganization is much faster than the bare diffusion arising from sources and sinks of uncharged species.

During active pumping, the long-ranged ionic reorganization facilitates rapid electrochemical signal propagation along the membrane. 
Specifically, we find that the transmembrane potential $\Delta V^{\rm M}$ exhibits three distinct scaling regimes: a near-field region where $\Delta V^{\rm M}$ is uniform, an intermediate region where ${\Delta V^{\rm M}(r) \propto 1/r}$ and a far-field region where ${\Delta V^{\rm M}(r) \propto 1/r^3}$, with $r$ being the radial distance from the transporter. As long as there is active ionic current, the intermediate region expands with a constant speed that is enhanced by a factor ${\bar{\Gamma} \bar{\delta}^{M} / 2 \approx 40}$ where $\bar{\Gamma}$ quantifies the dielectric mismatch between the lipid membrane and the surrounding electrolyte, and $\bar{\delta}^{M}$ is the dimensionless membrane thickness relative to the width of the diffuse charge layer.

The paper is organized as follows. In section \ref{sec:Model}, we describe the model, non-dimensionalize the governing PNP equations and boundary conditions, and outline the numerical simulation strategy using the finite element method (FEM). In section \ref{sec:uniform} we analyze our model in a pedagogical setting where in-plane variations can be neglected, corresponding to spatially uniform pumping across a large membrane patch. This simplification results in a 1D model that is analytically tractable, and we show that it recapitulates the classical equivalent circuit picture when in a steady state. Then, in section \ref{sec:local}, we analyze the spatiotemporal dynamics during localized transport from a single ion pump. We first present the FEM results, which describe the phenomenology of charge reorganization and in-plane scaling regimes. Then, we present a theoretical analysis that reveals the physical mechanisms underlying these results and identifies the crucial role of long-ranged electrostatic forces and the dielectric mismatch between the lipid membrane and the surrounding medium in the in-plane propagation of the electrochemical signal. Implications for electrical excitability of neuronal membranes involving multiple ion transporters acting in concert are discussed.

\section{Problem Description}\label{sec:Model}

\subsection{System and Governing Equations}

Consider a planar, uncharged lipid membrane (M) of thickness~${\delta^{\rm M}\sim4 \,{\rm nm}}$ separating two semi-infinite domains ${\alpha\in\{\rm I,II\}}$ each containing a dilute electrolyte solution (Fig.~\ref{fig1_schematic_descriptions}). The electrolyte solutions have permittivity ${\epsilon\sim80 \epsilon_{0}}$, where $\epsilon_{0}$ is the vacuum permittivity. The permittivity of the lipid membrane~($\epsilon^{\rm M}$) is much smaller, with ${\epsilon^{\rm M} \sim 4 \epsilon_{0}}$. The ratio of the membrane and solution permittivity is represented by ${\bar{\Gamma}\equiv\epsilon/\epsilon^{\rm M} \sim 20}$, which quantifies the dielectric mismatch. 

In the dilute limit of concentrated electrolyte theory \cite{fong2020transport} and in the absence of advection, the spatiotemporal dynamics of the electrolytes are described by the PNP equations \cite{Nernst1888,Nernst1889,Planck1890,Bazant04},
\begin{subequations}

\begin{align}
    \frac{\partial C^\alpha_i}{\partial t} &\eq -\bnabla \cdot \bj_i^\alpha \ , \label{eq:dim_eq_C} \\
    \bj_i^\alpha 
    &\eq
    -D_i^\alpha
    \left(
        \bnabla C_i^\alpha
            + \frac{{\rm z}_{i}{\rm e}}{k_{\mathrm{B}}T} C_i^\alpha \bnabla\phi^\alpha
    \right) \ , \label{eq:dim_eq_ionFlux} \\
    \bnabla\cdot\bD^\alpha
    &\eq
    \rho^\alpha \eq \sum_{i}{\rm z}_{i}{\rm e} C_{i}^\alpha \ , \label{eq:dim_eq_poisson}\\
    \bD^\alpha &\eq -\epsilon\bnabla\phi^\alpha \ \label{eq:dim_eq_linDielec},
\end{align}
where $C^\alpha_i (\bx,t)$ is the concentration of ionic species ${i\in\{1,\ldots,N\}}$ at location $\bx$ and time $t$ within domain $\alpha$. For each ionic species, Eq.~\eqref{eq:dim_eq_C} is the mass balance and Eq.~\eqref{eq:dim_eq_ionFlux} is the constitutive relation for the flux $\bj_i^\alpha$ with terms corresponding to Fickian diffusion and electromigration. Here, $D_i^\alpha$ is the diffusivity, ${\rm z}_i$ is the ion valence, $k_{\mathrm{B}}T$ is the thermal energy scale, and ${\rm e}$ is the fundamental charge. Equation~\eqref{eq:dim_eq_poisson} is Gauss's law, where $\bD^\alpha(\bx,t)$ is the electric displacement field and $\rho^\alpha(\bx,t)$ is the ionic charge density. Finally, Eq.~\eqref{eq:dim_eq_linDielec} relates the electric displacement field $\bD^\alpha(\bx,t)$ to the gradient of the electric potential $\phi^\alpha(\bx,t)$.  Note that under the electrostatic approximation, the electric field ${\mathbf{E}^\alpha(\bx,t) = - \bnabla\phi^\alpha(\bx, t)}$ \cite{jackson1999classical,Kovetz2000}.
  
Lipid membranes solvate ions poorly due to the low dielectric permittivity in their interior \cite{volkov2008energetics, nymeyer2008method}. Consequently, the concentrations of ionic species are extremely low inside the lipid membrane and Gauss's law can be written as
\begin{align}
    \bnabla\cdot\bD^{\mathrm{M}} & \ \approx\  0 \ , \label{eq:dim_eq_poissonM}\\
    \bD^{\mathrm{M}} &\eq -\epsilon^{\mathrm{M}}\bnabla\phi^{\mathrm{M}} \ ,\label{eq:dim_eq_linDielecM}
\end{align}
\end{subequations}
where $\phi^{\rm M}$ is the membrane electric potential. Equations~\eqref{eq:dim_eq_C}-\eqref{eq:dim_eq_linDielecM} govern the spatiotemporal dynamics of our system. In this work, we consider the simplest scenario corresponding to a symmetric monovalent electrolyte (${i\in\{+,-\}}$ and ${{\rm z}_{+}=-{\rm z}_{-}=1}$) with equal diffusivity for both ions (${D_i^\alpha=D\ \forall \; i, \alpha}$). We assume constant $\epsilon$ and $\epsilon^{\rm M}$, in which case Eqs.~\eqref{eq:dim_eq_poisson}-\eqref{eq:dim_eq_linDielec} and Eqs.~\eqref{eq:dim_eq_poissonM}-\eqref{eq:dim_eq_linDielecM}, respectively, can be combined to yield Poisson's equation in each domain.

\subsection{Initial and Boundary Conditions}
\label{sec:Model_icbc}

Initially, the electrolyte solutions are homogeneous and symmetric, i.e.  ${C_{\pm}^\alpha(\bx,t=0)=C^0}$ (constant). Starting at $t=0$, ions are pumped across the membrane. Without loss of generality, we consider the selective pumping of cations, as is typically the case involving ion channels and pumps in many neuronal membranes \cite{hille1992,SKOU1957394}. This amounts to 
\begin{subequations}
\begin{alignat}{3}
    \bj_+^{\rm I} \cdot \bn^{\rm I} &\eq j^{\rm o}(x, y, t) \quad &&\textrm{on }S^{\rm I} \ , \label{eq:dim_bc_imposedFlux1}\\
    \bj_+^{\rm II} \cdot \bn^{\rm II} &\eq -j^{\rm o}(x, y, t) \quad &&\textrm{on }S^{\rm II} \ , \label{eq:dim_bc_imposedFlux2}\\
    \bj_-^\alpha\cdot\bn^\alpha &\eq 0\quad &&\textrm{on }S^{\alpha}\ ,\quad \alpha\in\{\rm I,II\} \ , \label{eq:dim_bc_anionNoFlux}
\end{alignat}
where $S^{\alpha}$ represents the membrane-fluid interface for domain $\alpha$, $\bn^\alpha$ is the outward normal pointing into the membrane with $\bn^{\rm I}=-\bn^{\rm II}=\bbe_z$ (Fig.~\ref{fig1_schematic_descriptions}), and $j^{\rm o}$ is the imposed cation flux from domain $\rm I$ to $\rm II$. The fluxes at $S^{\rm I}$ and $S^{\rm II}$ are equal in order to ensure global electroneutrality. Electrolyte solutions remain unperturbed by the pumping far from the membrane: $|\bx|\!\rightarrow\!\infty$, $C_{\pm}^{\alpha}(\bx,t)\!\rightarrow\!C^{0}$ and $\bD ^\alpha(\bx,t)\!\rightarrow\!\te{0}$.

Electrostatics requires that the electric potential is continuous at the membrane-fluid interface, i.e. 
\begin{equation}
    \phi^\alpha \eq \phi^{\mathrm{M}} \quad \textrm{on }S^{\alpha}\ ,\quad \alpha\in\{\rm I,II\}  \ .\label{eq:dim_bc_phiConti}
\end{equation}
In this work, the membrane does not carry any surface charge, so the normal component of the electric displacement field is continuous at the membrane-fluid interface \cite{Kovetz2000}, i.e.
\begin{equation}
    \left(\bD^{\mathrm{M}}-\bD^\alpha\right)\cdot\bn^\alpha \eq 0 \quad \textrm{on }S^{\alpha}\ ,\quad \alpha\in\{\rm I,II\}  \ .\label{eq:dim_bc_DConti}
\end{equation}
\end{subequations}

\subsection{Non-dimensionalization}
We nondimensionalize the governing equations and boundary conditions using the following scales for length, time, and potential,
\begin{equation}
    \lambda_{\mathrm{D}} = \sqrt{\frac{\epsilon k_{\rm B}T}{2C^0 \mathrm{e}^2}}\ , \;\;\;\;\; 
    \tau_{\mathrm{D}} = \frac{\lambda_{\mathrm{D}}^2}{D}\ , \;\;\;\;\;  \phi_{\mathrm{T}} = \frac{k_{\rm B}T}{e}\ .
    \label{eq:def_NDscales}
\end{equation}
Here, $\lambda_{\mathrm{D}}$ is the Debye screening length that represents the thickness of the diffuse charge layer \cite{debye1923,Gouy1910, Bazant04}, $\tau_{\mathrm{D}}$ is the Debye time corresponding to the diffusion time scale of the diffuse layer \cite{Bazant04}, and $\phi_{\mathrm{T}}$ is the thermal voltage. In biological settings with a salt concentration ${C^0\sim 150\,{\rm mM}}$ \cite{lodish2000molecular} and ${D\sim 1 {\rm nm^2/ns}}$ \cite{vanysek2015}, their characteristic values are ${\lambda_{\mathrm{D}} \sim 1 \,{\rm nm}}$, ${\tau_{\mathrm{D}} \sim 1 \,{\rm ns}}$, and ${\phi_{\rm T} \sim 25\,{\rm mV}}$. 

Throughout this paper, we use $\overline{\left( \cdot \right)}\,$ to denote dimensionless variables. Given a symmetric monovalent electrolyte, the system of equations~\eqref{eq:dim_eq_C}-\eqref{eq:dim_eq_linDielec} can also be analyzed, without loss of generality, in terms of the charge density and the overall salt concentration. To that end, introducing the dimensionless charge density ${\bar{\rho}^\alpha\!\equiv\! (C_{+}^\alpha-C_{-}^\alpha)/(2C^0)}$, the dimensionless disturbance in salt concentration ${{\bar{C}^\alpha\!\equiv\!(C_{+}^\alpha + C_{-}^\alpha - 2C^0)/(2C^0)}}$ and the dimensionless electric potential ${\bar{\phi}^\alpha\!\equiv\!\phi^\alpha/\phi_{\rm T}}$, Eqs.~\eqref{eq:dim_eq_C}-\eqref{eq:dim_eq_linDielec} can be rewritten as,
\begin{subequations}
\begin{align}
    \frac{\partial \bar{\rho}^\alpha}{\partial \bar{t}} + \bar{\bnabla} \cdot \bar{\bj}_{\rho}^{\alpha} 
    &\eq 0 \ , 
     \label{eq:nondim_eq_rho}\\
    \frac{\partial \bar{C}^\alpha}{\partial \bar{t}} + \ \bar{\bnabla} \cdot  \bar{\bj}_{C}^{\alpha} 
    &\eq 0 \ , \label{eq:nondim_eq_C}\\
    \quad \ \ -\bar{\nabla}^2 \bar{\phi}^\alpha
    &\eq \bar{\rho}^\alpha \ ,\label{eq:nondim_eq_poisson}
\end{align}
where $\bar{\bnabla}\!\equiv\!\partial/\partial \bar{\bx}$, $\bar{\bx}\!\equiv\!\bx/\lambda_{\rm D}$, and $\bar{t}\!\equiv\!t/\tau_{\rm D}$. The charge and the salt fluxes are given by
\begin{align}
\bar{\bj}_{\rho}^{\alpha}
    &\eq
    - \bar{\bnabla} \bar{\rho}^\alpha -\left(1 + \bar{C}^\alpha  \right) \bar{\bnabla} \bar{\phi}^\alpha \ ,
     \label{eq:nondim_eq_rhoFlux} \\
    \bar{\bj}_{C}^{\alpha}
    &\eq 
    -\bar{\bnabla} \bar{C}^\alpha -\bar{\rho}^\alpha \bar{\bnabla} \bar{\phi}^\alpha \ .\label{eq:nondim_eq_CFlux}
\end{align}
Also, from equations~\eqref{eq:dim_eq_poissonM} and~\eqref{eq:dim_eq_linDielecM}, the electrostatics inside the membrane is described by
\begin{equation}
    \bar{\nabla}^{2} \bar{\phi}^{\rm M} \eq 0 \ . \label{eq:nondim_eq_poissonM}
\end{equation}
\end{subequations}

The initial and boundary conditions can be nondimensionalized as well. Initially, electrolyte solutions are homogeneous: $\bar{\rho}^\alpha \!=\! 0$, $\bar{C}^\alpha \!=\! 0$, $\bar{\phi}^\alpha \!=\!0$, and $\bar{\phi}^{\rm M} \!=\!0$. Dimensionless boundary conditions for the charge and salt balance fluxes at the membrane-fluid interfaces based on Eqs.~\eqref{eq:dim_bc_imposedFlux1}-\eqref{eq:dim_bc_anionNoFlux} are 
\begin{subequations}
\begin{alignat}{4}
    &\bar{\bj}_{\rho}^{\rm I} \cdot \bn^{\rm I} &&\eq \bar{\bj}_{C}^{\rm I} \cdot \bn^{\rm I} && \eq \bar{j}^{\rm o}(\bar{x},\bar{y},\bar{t}) \quad &&\textrm{on }\bar{S}^{\rm I} \ , \label{eq:nondim_bc_imposedFlux1}\\
    &\bar{\bj}_{\rho}^{\rm II} \cdot \bn^{\rm II} &&\eq \bar{\bj}_{C}^{\rm II} \cdot \bn^{\rm II} && \eq -\bar{j}^{\rm o}(\bar{x},\bar{y},\bar{t}) \quad  &&\textrm{on }\bar{S}^{\rm II} \ ,\label{eq:nondim_bc_imposedFlux2}
\end{alignat}
where $\bar{j}^{\rm o}\!\equiv\!j^{\rm o} \tau_{\rm D}/ (2C^{0} \lambda_{\rm D})$. We nondimensionalize the continuity of the electric potential~\eqref{eq:dim_bc_phiConti} and displacement field~\eqref{eq:dim_bc_DConti} conditions as
\begin{alignat}{2}
    \bar{\phi}^\alpha &\eq \bar{\phi}^{\rm M} \quad &&\textrm{on }\bar{S}^{\alpha}\ ,\  \alpha\in\{\rm I,II\} \ , \label{eq:nondim_bc_phiConti}\\
    \bar{\bnabla}\,\bar{\phi}^\alpha\cdot\bn^\alpha &\eq \frac{1}{\bar{\Gamma}}\bar{\bnabla}\,\bar{\phi}^{\rm M}\cdot\bn^\alpha \quad &&\textrm{on }\bar{S}^{\alpha}\ ,\ \alpha\in\{\rm I,II\} \ ,\label{eq:nondim_bc_DConti}
\end{alignat}
\end{subequations}
where $\bar{\Gamma}$ is the dielectric ratio introduced earlier. 
Lastly, as ${|\bx|\to\infty}$, the boundary conditions we impose are ${\bar{\rho}^\alpha\to 0}$, ${\bar{C}^\alpha\to 0}$, and ${\bar{\bD}^\alpha\to \te{0}}$.

\subsection{Numerical Solution Strategy}
In what follows, the above system of coupled partial differential equations is analyzed in two scenarios as shown in Fig.~\ref{fig2_schematic_uniform}: spatially uniform pumping (Sec.~\ref{sec:uniform}) and spatially localized pumping of cations (Sec.~\ref{sec:local}). Solutions to the full non-linear equations are obtained numerically using the finite element method (FEM)~\cite{hughes2012finite,reddy2015introduction,papadopoulos2015fem}. We employ the packages \texttt{FEniCSx}~\cite{Scroggs22a,Scroggs22b,Alnaes14} for FEM implementation and \texttt{Gmsh}~\cite{Geuzaine09} for mesh construction. For the uniform pumping model, we use one-dimensional linear elements, and for the localized pumping model, we use bilinear quadrilateral elements with jump boundary conditions enforced weakly.

As shown below, the localized pumping problem contains multiple length and time scales. The different length scales include the pore radius of the transporter (${\sim 2 \, \rm{\mathring{A}}}$), the Debye length (${\sim 1\, \rm{nm}}$), and the induced electric fields (${\rm{nm}-\rm{mm}}$). The time scales include the Debye time (${\sim 1\, \rm{ns}}$), the pump open times (${\si{\micro s}-\rm{ms}}$), and the time scale associated with the spatial propagation of the electrical fields ($\rm{ms}$). Resolving behaviors over these length and time scales requires careful design of the meshing and time stepping strategies. Complete details are described in \jbfern{SM Secs.~I and II} (also see Movie~\href{https://youtube.com/shorts/QFAwuHB3ACk}{S1}).

\section{Spatially Uniform Ion Pumping: 1D Model}
\label{sec:uniform}

\begin{figure}[t]
\centering
\includegraphics[width=\linewidth]{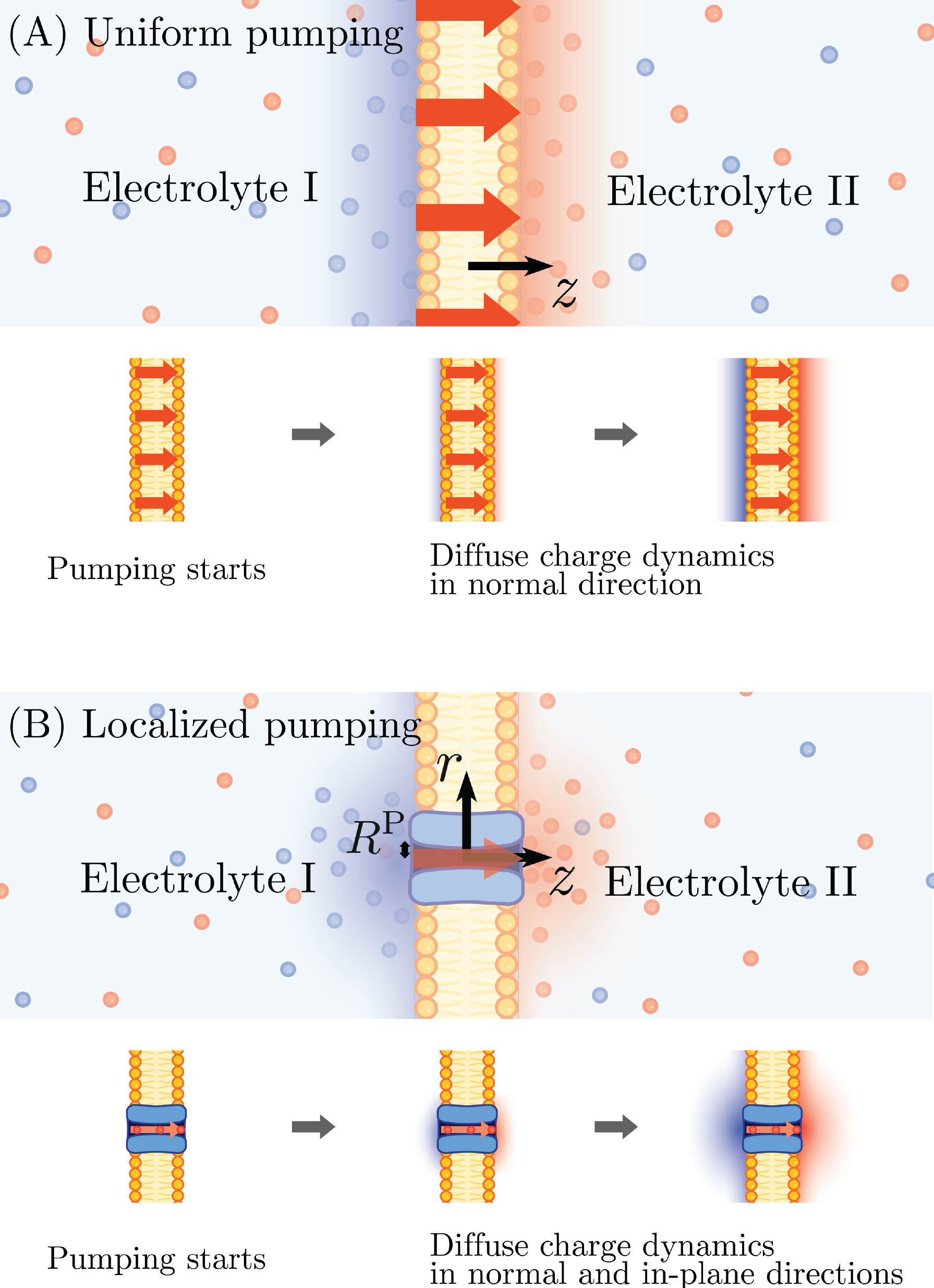}
    \caption{
    Schematics of the two modeling scenarios considered in this work. (A) Uniform pumping: cations are pumped uniformly over the membrane surface, resulting in an effective 1D spatial model where only variations along the normal direction $z$ are relevant. Blue and red shades represent positive and negative charge densities, respectively. The lower panel shows the transient build-up of the diffuse charge layers in the normal direction. (B) Localized pumping: cations are transported in a spatially localized fashion through a pump modeled as a membrane patch of radius $R^{\rm P}$. The lower panel indicates the transient profiles of the diffuse charge vary along both the axial ($z$) and radial ($r$) directions.
    }
\label{fig2_schematic_uniform}
\end{figure}

For pedagogical purposes, we first analyze our model in a 1D setting where the cation flux is uniform over the plane of the membrane (Fig.~\ref{fig2_schematic_uniform}(A)). While biologically unrealistic, this simplification is often implicit in classical equivalent circuit models \cite{cole1938,hodgkin1946,rall1962, dayan2001}, which altogether neglect diffuse charge dynamics. However, due to the ion-selective nature of the pumping in our model, diffuse charge layers of thickness ${\sim\lambda_{\rm D}}$ are formed near the membrane-solution interface. Accordingly, we solve the spatiotemporal dynamics of this 1D model numerically and analytically. We find that the membrane and the resulting diffuse charge layers resemble the equivalent circuit picture assumed in classical models, albeit only at a steady state. 

The uniform pumping of cations amounts to imposing boundary conditions~\eqref{eq:nondim_bc_imposedFlux1} and~\eqref{eq:nondim_bc_imposedFlux2} of the form
\begin{equation}
    \bar{j}^{\rm o}\left(\bar{x}, \bar{y}, \bar{t}\right)
    \eq
    \begin{cases}
        \ \bar{j}^{\rm o}_0 \ \mbox{(const.)} & \quad 0 \leq \bar{t} < \bar{\tau}^{\rm o}\\
        \ 0 & \quad \mbox{otherwise}
    \end{cases}
    , \label{eq:def_uniformFlux}
\end{equation}
where $\bar{\tau}^{\rm o}$ and $\bar{j}_0^{\rm o}$ represent the duration and the rate of pumping, respectively. We consider cases where the pump open times are much larger than the Debye time, i.e. ${\bar{\tau}^{\rm o}\gg 1}$.
It is instructive to determine the transmembrane potential difference, ${{\Delta\bar{V}^{\rm M}} \equiv \left. \bar{\phi}^{\rm II} \right|_{S^{\rm II}} -  \left. \bar{\phi}^{\rm I} \right|_{S^{\rm I}}}$, as a function of the total pumped charge per area, ${\bar{Q}(\bar{t})= \bar{j}^{\rm o}_0 \bar{t}}$.
From Gauss's law, for this 1D problem, ${\Delta\bar{V}^{\rm M}}$ and $\bar{Q}$ obey a linear relation: ${\Delta\bar{V}^{\rm M}=\bar{\Gamma}\bar{\delta}^{\mathrm{M}}\bar{Q}}$ \jbfern{(SM Sec.~III.1)}.
In physiological settings for neuronal signaling, ${\Delta\bar{V}^{\rm M}\sim\mathcal{O}(1)}$ (${\sim25\,\mathrm{mV}}$) \cite{hodgkin1939action, hille1992}, ${\bar{\Gamma}\sim 20}$, and ${\bar{\delta}^{\mathrm{M}}\sim 4}$, and therefore only a small amount of charge transport ${\bar{Q}\sim1/80 \ll1}$ is required. 
This also motivates a perturbation analysis in the limit of small $\bar{j}^{\rm o}_0$ \jbfern{(SM Sec.~III.2)}. In what follows, we compare the leading order solutions from the perturbation analysis with the FEM solutions of the full PNP equations. We first discuss the behavior for ${\bar{t} \leq \bar{\tau}^{\rm o}}$ (\emph{charging dynamics}), followed by ${\bar{t} > \bar{\tau}^{\rm o}}$ (\emph{relaxation dynamics}). For complete analytical solutions and their derivations, see \jbfern{SM Sec.~{III.3}}.

\begin{figure}
    \centering
    \includegraphics[width=\linewidth]{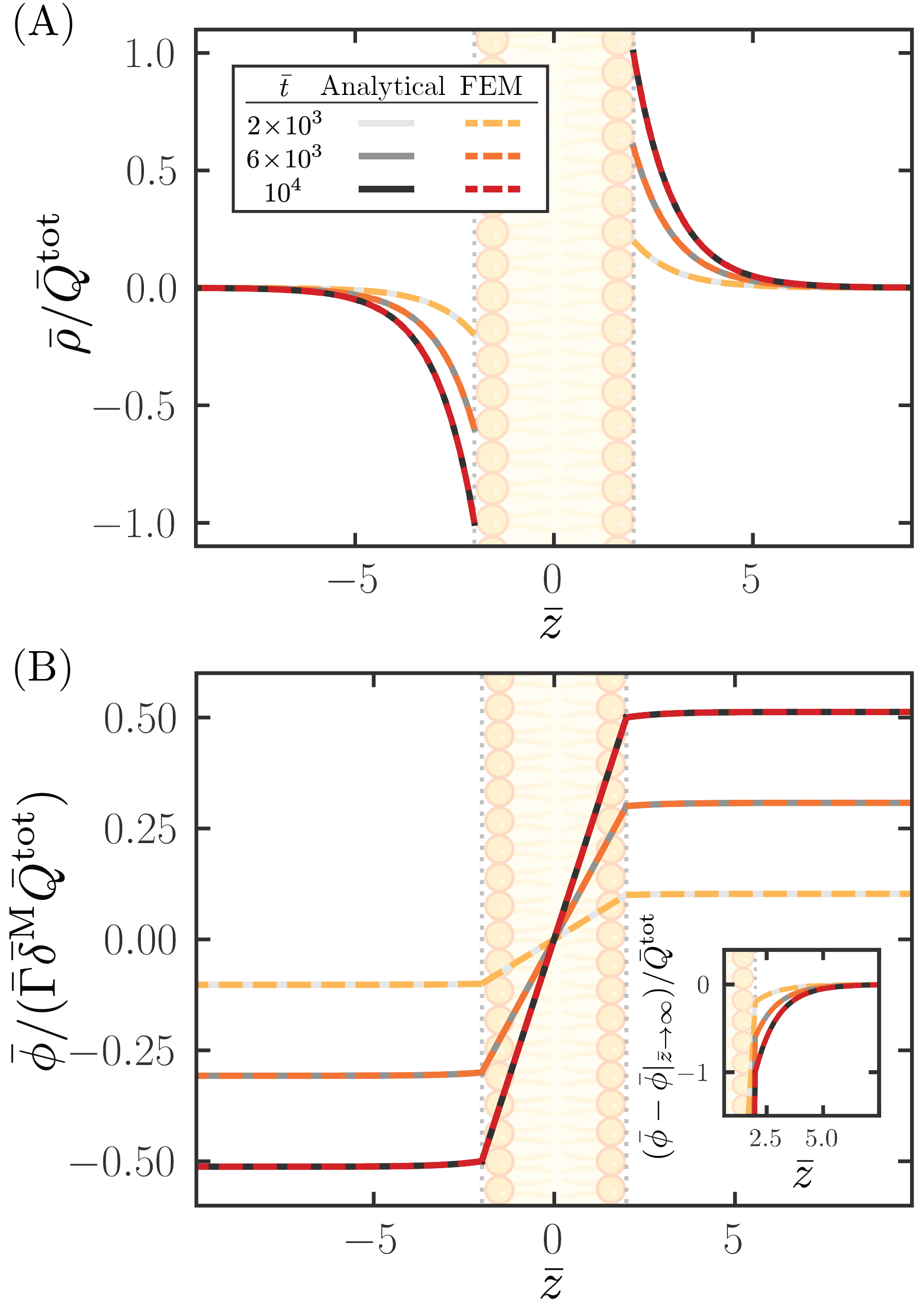}
    \caption{Transient dynamics of (A) the charge density and (B) the electric potential during uniform pumping in the charging regime. In both panels, analytical solutions (solid lines) at different times $\bar{t}$ are compared with FEM solutions (dashed lines). Charge densities are scaled by the total amount of transported charges per area ${\bar{Q}^{\rm tot}\ (=\bar{j}^{\rm o}_0\bar{\tau}^{\rm o})}$ and electric potentials are scaled by the potential drop across the membrane (${=\bar{\Gamma}\bar{\delta}^{\rm M}\bar{Q}^{\rm tot}}$). The inset in (B) shows a magnified view of electric potential variation in the diffuse charge layer. Simulation parameters used: ${\bar{\Gamma}=20}$, ${\bar{\delta}^{\rm M}=4}$, ${\bar{j}^{\rm o}_0=1.25\times 10^{-6}}$, and ${\bar{\tau}^{\rm o}=10^{4}}$.
    }
\label{fig3_uniform_pumping}
\end{figure}

\subsection{Charging Dynamics ($0 \leq \bar{t} \leq \bar{\tau}^{\rm o}$)}

Figure~\ref{fig3_uniform_pumping} shows the time-dependent charge density and electric potential during pumping. The leading-order perturbation equations for the charge density, potential, and salt concentration disturbance form a linear system that is odd with respect to the normal coordinate $\bar{z}$. This enables analytical solutions, by considering only one of the two domains. For the charge density and potential, the asymptotic solutions for domain II  when ${\bar{t}\gg1}$ are
\begin{align}
    \bar{\rho}^{\rm II}(\bar{z}', \bar{t}) & \eq \bar{j}_0^{\rm o}\left[\bar{t}+\frac{1}{2}(1-\bar{z}')\right]e^{-\bar{z}'} + \mathcal{O}\left(\sqrt{\bar{t}}e^{-\bar{t}}\right) \ , \label{eq:charge_rho} \\
    \bar{\phi}^{\rm II}(\bar{z}', \bar{t}) & \eq \bar{j}_0^{\rm o}\bigg[\frac{\bar{\Gamma}\bar{\delta}^{\mathrm{M}}}{2}\bar{t}+\left(\bar{t}-\frac{1}{2}\right)\left(1-e^{-\bar{z}'}\right)  \notag \ \\
    & \qquad\qquad\qquad\quad + \bar{z}'e^{-\bar{z}'}\bigg] +\mathcal{O}\left(\sqrt{\bar{t}}e^{-\bar{t}}\right) \ , \label{eq:charge_phi}
\end{align}
where ${\bar{z}'=\bar{z}-\bar{\delta}^{\rm M}/2}$ is the distance from the membrane surface. In the 1D case, the membrane potential $\bar{\phi}^{\rm M}$ varies linearly in the membrane. The salt concentration $\bar{C}$ spreads diffusively in a self-similar fashion, akin to an uncharged solute \jbfern{(SM Sec.~III.4)}. Figure~\ref{fig3_uniform_pumping} shows the FEM and analytical solutions for $\bar{\rho}$ and $\bar{\phi}$ agree well. 

Figure~\ref{fig3_uniform_pumping}(A) shows that pumped charges result in a diffuse charge layer of net positive charge in domain II, localized within ${\sim5}$ Debye lengths ($\lambda_\mathrm{D}$) near the membrane-fluid interface. The formation of this diffuse charge layer can be understood by the following physical argument: the net result of removing cations from domain I is to impose an effective negative surface charge 
on the system comprising domain II and the insulating membrane. This scenario is similar to the classical problem of electrolyte screening near a flat, blocking electrode considered by Gouy \cite{Gouy1910}  and Chapman \cite{chapman1913li}. Accordingly, the diffuse charge layer in domain II, which has a net positive charge, arises in order to screen the electric field due to this negative surface charge.
Simultaneously, due to global electroneutrality and symmetry, the system also develops a diffuse charge layer of equal and opposite charge in domain I at the membrane-fluid interface. Thus, the continuous pumping of cations amounts to charging the membrane as a capacitor with charges $-\bar{Q}(\bar{t})$ and $+\bar{Q}(\bar{t})$ on either side.

Figure~\ref{fig3_uniform_pumping}(B) and its inset show that the drop in electric potential along $\bar{z}$ primarily occurs inside the membrane rather than in the diffuse layers. 
The diffuse layer potential ${\Delta\bar{V}^{\rm DL}\equiv \bar{\phi}|_{\bar{z}'\to\infty} - \bar{\phi}|_{\bar{z}'=0}}$, which from Eq.~\eqref{eq:charge_phi}, is ${\Delta\bar{V}^{\rm DL} \approx \bar{Q}(\bar{t})  \ll \Delta\bar{V}^{\rm M}  = \bar{\Gamma}\bar{\delta}^{\mathrm{M}}\bar{Q}(\bar{t}) }$.
Thus, the low dielectric permittivity of the lipid membrane compared to the surrounding medium and its large thickness compared to the Debye layer contribute to the dominant potential drop inside the membrane. In its dimensional form, the transmembrane potential ${\Delta{V}^{\rm M}= {Q}(t)/C_{\rm M}}$, which is again consistent with the idea of the membrane as a capacitor with capacitance per unit area ${C_{\rm M} = \epsilon^{\rm M}/\delta^{\rm M}}$. 

For a constant pumping rate $\bar{j}_0^{\rm o}$, $\Delta \bar{V}^{\rm M}$ linearly increases with time (Fig.~\ref{fig3_uniform_pumping}(B)). However, this cannot continue indefinitely as we find that for any value of $\bar{j}^{\rm o}_0$, there is always a \emph{maximum open time} $\bar{\tau}^{\rm o}_{\rm max}\propto{(\bar{j}^{{\rm o}}_0)^{-2}}$ beyond which concentrations near the pump in domain I become negative \jbfern{(SM Sec.~III.5)}. Consequently, in the 1D setting,  pumping cannot be indefinite even for arbitrarily small values of the pumping rate $\bar{j}^{\rm o}_0$. This, as we shall see, is not the case for localized pumping in a 3D system.

\subsection{Relaxation Dynamics ($\bar{t} > \bar{\tau}^{\rm o}$)}

We now discuss the relaxation dynamics of the system towards a steady state, once the cationic pumping is deactivated.
Again, due to the linearity of the equations, the relaxation dynamics to the leading order can be obtained by subtracting a time-shifted copy of the charging solution from itself \jbfern{(SM Sec.~III.6)}. For times ${\bar{t}\gg\bar{\tau}^{\rm o}}$, we find that 
\begin{align}
    \bar{\rho}^{\rm II}(\bar{z}', \bar{t}) & \eq \bar{Q}^{\rm tot}e^{-\bar{z}'} + \mathcal{O}\left(\sqrt{\bar{t}}e^{-\bar{t}}\right) \ , \label{eq:relax_steady_rho} \\
    \bar{\phi}^{\rm II}(\bar{z}', \bar{t}) & \eq
    \bar{Q}^{\rm tot}\left(\bar{\Gamma}\frac{\bar{\delta}^{\mathrm{M}}}{2}+1-e^{-\bar{z}'}\right) \notag \ \\
    & \qquad\qquad\qquad\quad  +
    \mathcal{O}\left(\sqrt{\bar{t}}e^{-\bar{t}}\right), \label{eq:relax_steady_phi}
\end{align}
where ${\bar{z}'=\bar{z}-\bar{\delta}^{\rm M}/2}$ is the distance from the membrane surface and ${\bar{Q}^{\rm tot}=\bar{j}^{\rm o}_0\bar{\tau}^{\rm o}}$ is the total pumped charge per unit area.

Equations ~\eqref{eq:relax_steady_rho} and \eqref{eq:relax_steady_phi} show that both $\bar{\rho}$ and $\bar{\phi}$ relax to their steady-state profiles with a time scale ${\tau_D \sim 1 \; \text{ns}}$, which is much smaller than the value of $\bar{\tau}^{\rm o}$ considered here, as well as the known time scales of pump or channel activity ${\sim 1  \, \rm{ms}}$ \cite{hille1992}, or macroscopic processes such as action potentials \cite{hodgkin1939action,hille1992}. The steady-state solutions for $\bar{\rho}$ and $\bar{\phi}$ also show that in either case the spatial variation is limited to the diffuse charge layers of length scale $\lambda_{\rm D}$.

\begin{figure}[t]
    \centering
    \includegraphics[width=\linewidth]{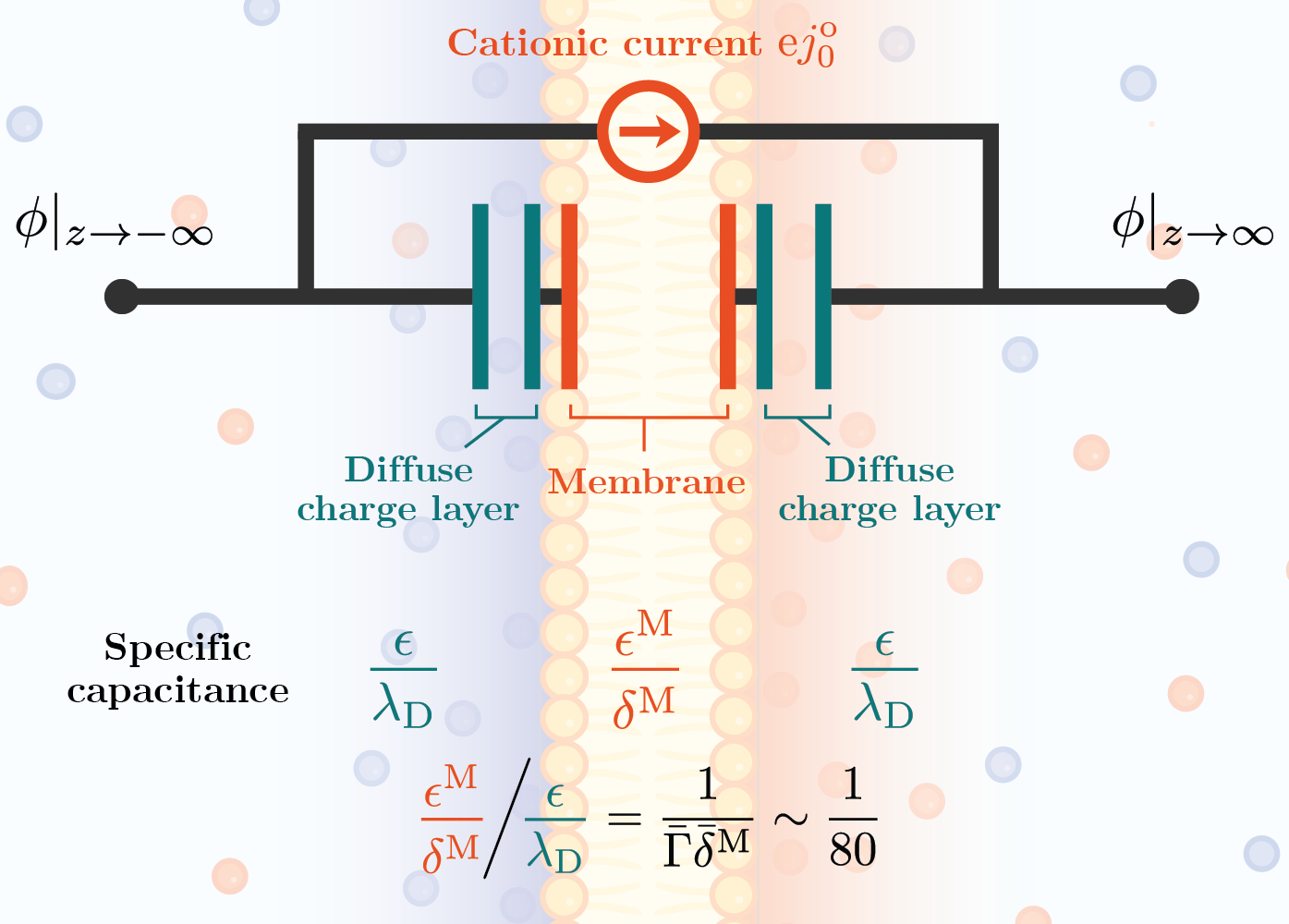}
    \caption{The equivalent circuit picture for the uniform pumping model at steady state. The membrane and the surrounding diffuse charge layers are represented as capacitors in series. The capacitance per unit area of the membrane is $C_{\rm M} = \epsilon^{\rm M} / \delta^{\rm M}$ and that of each diffuse layer is $C_{\rm D} = \epsilon / \lambda_{\rm D}$. As capacitors in series add harmonically, the membrane contribution to the total capacitance $C$ dominates the contributions of the diffuse charge layers, i.e. $C = \left( C_{\rm M}^{-1} + 2 C_{\rm D}^{-1} \right)^{-1} \approx C_{\rm M}$.}
    \label{fig4_capacitor}
\end{figure}

\begin{figure*}
    \centering
    \includegraphics[width=\textwidth]{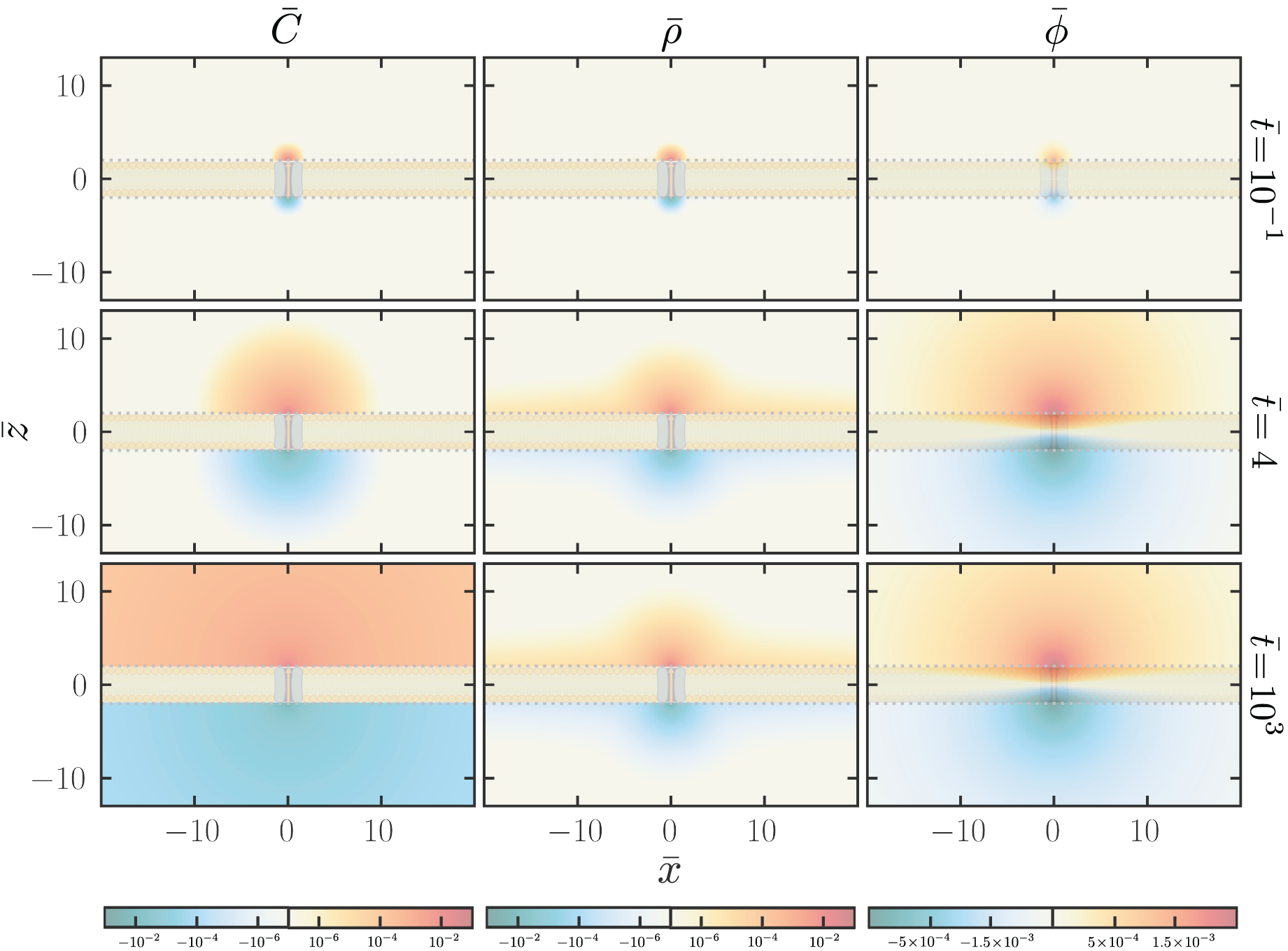}
    \caption{Heatmaps of the FEM solutions in $(\bar{x}, \bar{z})$-plane about the center of channel for the salt concentration (left), charge density (center), and electric potential (right) near the pump at times ${\bar{t}=10^{-1},\ 4}$, and $10^3$ (top to bottom row). Note that the system is axisymmetric around the vertical $\bar{z}$-axis.
    FEM simulations of the full PNP equations are performed using parameters ${\bar{\Gamma}=20}$, ${\bar{\delta}^{\rm M}=4}$, ${\bar{j}^{\rm o}_0=4}$, and ${\bar{R}^{\rm P}=0.1}$.  The FEM solutions demonstrate that the charge density and electric potential reach a steady-state profile near the pump within ${\bar{t} = \mathcal{O}(1)}$ during active pumping. In contrast, the salt concentration evolves in a diffusive manner with spherical symmetry, akin to an uncharged solute. All data are presented in logarithmic scales (color bar, bottom).  
    }
\label{fig5_contour_plots}
\end{figure*}

\subsection{Conclusions}
We summarize key features of the uniform pumping model before considering the case of localized pumping in the next section. \emph{First}, we find that the leading order perturbation analysis of the PNP equations accurately recovers the dynamics of $\bar{\rho}$, $\bar{\phi}$ and $\bar{C}$ in biologically relevant settings. \emph{Second}, the dynamics of charging and relaxation are governed by the Debye length $\lambda_D$ and the Debye time $\tau_D$. The system reaches a well-defined steady state within $\mathcal{O}(\tau_D)$ once the pumping is stopped. \emph{Third}, for a given pumping rate $\bar{j}^{{\rm o}}_0$, there is a maximal pumping time above which ionic concentrations can become negative. \emph{Fourth}, the dielectric mismatch between the membrane and electrolyte enhances the transmembrane potential, as seen by the linear relationship between $\Delta\bar{V}^{\rm M}$ and $\bar{\Gamma}$. \emph{Finally}, the transmembrane potential $\Delta\bar{V}^{\rm M}$ varies linearly with $\bar{Q}$, suggesting that in the 1D model, the lipid membrane can be conveniently approximated as a capacitor.

At steady state, the lipid membrane and the surrounding diffuse layers on either side behave like parallel-plate capacitors in series, as shown in Fig.~\ref{fig4_capacitor}.
By contrast, there does not appear to be an equivalent circuit that captures the transient nature of the dynamics (${\bar{t} < 1}$), as it is not obvious how to represent the resistances to ionic motion in the diffuse layers as circuit elements. Nevertheless, the system reaches a quasi-steady state on time scales larger than $\tau_{\rm D}$, and on these time scales the diffuse layer potential difference can be approximated as ${\Delta\bar{V}^{\rm DL}\approx\bar{Q}(t)}$.
Upon dimensionalization, this yields ${\Delta{V}^{\rm DL}\approx [1/(\epsilon/\lambda_{\rm D})] {Q}(t)}$, which is captured by the diffuse layer capacitance ${C_{\rm D} = \epsilon/\lambda_{\rm D}}$ in Fig.~\ref{fig4_capacitor}. 
The membrane capacitance is smaller than the electrolyte capacitance by a factor of ${\bar{\Gamma} \bar{\delta^{\rm M}} \approx 80}$, which is why ${\sim98\%}$ of the potential drop occurs inside the membrane consistent with Fig.~\ref{fig3_uniform_pumping}(B).
The equivalent capacitance of the full circuit is then a harmonic sum of the capacitances, which is dominated by the membrane capacitance (Fig.~\ref{fig4_capacitor}). 
This recovers the single capacitance form utilized in cable theoretic frameworks to model the membrane \cite{dayan2001}. 
While equivalent circuits may be suitable for describing one-dimensional systems, they typically fail to capture spatial heterogeneity in three-dimensional systems. This is especially true in our case where in-plane variations are significant, and may be necessary to understand electrochemical signal propagation along the membrane, which we now address.

\section{Spatially Localized Ion Pumping}
\label{sec:local}

Transmembrane ionic fluxes in biological settings occur through individual ion pumps and channel proteins, and are therefore spatially localized. Typical dimensions of the pores of ion channels and pumps are ${\sim 0.5 \, \mathrm{nm}}$ \cite{moldenhauer2016effective}. Furthermore, these proteins are sparsely distributed on the cell membrane; even in regions of high channel density such as axon hillocks and the nodes of Ranvier, the surface densities are estimated to be ${\sim 1000\,/ {\si{\micro m}}^2}$, implying a minimum distance of ${30\,\mathrm{nm}\ (\gg \lambda_{\rm D})}$ between neighboring transporters \cite{rosenbluth1976intramembranous}. Accordingly, we consider the problem of transport through a single pump and study the spatiotemporal aspects of charging and relaxation dynamics of the ionic concentrations and the electric potential.

We model the pump as a circular patch of radius $R^{\rm P}$ on an infinite planar membrane, as in Fig.~\ref{fig2_schematic_uniform}(B).  The system is axisymmetric about the transporter axis, with $\bar{z}$ representing the axial coordinate and $\bar{r}$ representing the radial coordinate.  The initial and boundary conditions are the same as introduced in Sec.~\ref{sec:Model_icbc}. The cation flux is now localized at the membrane-fluid interfaces, and is of the form  
\begin{equation}
    \bar{j}^{\rm o}(\bar{r}, \theta, \bar{t})
    \eq
    \begin{cases}
        \ \bar{j}^{\rm o}_0 \ \mbox{(const.)} & \bar{r} < \bar{R}^{\rm P}, \ 0 \leq \bar{t} < \bar{\tau}^{\rm o}\\
        \ 0 &  \mbox{otherwise}
    \end{cases}
    , \label{eq:def_uniformFlux}
\end{equation}
where ${\bar{R}^{\rm P}=R^{\rm P}/\lambda_{\rm D}}$ is the dimensionless radius of the patch, with $\bar{\tau}^{\rm o}$ and $\bar{j}^{\rm o}_0$ being the duration and rate of pumping, respectively. This system represents a classical source-sink problem toward understanding the spreading dynamics near biological membranes driven by a localized source and sink of charge. 
A non-zero cation flux breaks electroneutrality near the patch, creating a non-uniform electric field that reorganizes diffuse charges throughout the system. We now study the dynamics underlying this reorganization, beginning with the phenomenology from numerical simulations in Sec.~\ref{sec:Charging3D}, followed by a theoretical analysis in Secs.~\ref{sec:electrostatics}-\ref{sec:second-moments}. We analyze the relaxation dynamics towards the steady state upon pump deactivation in Sec.~\ref{sec:relaxation-3D}. Complete details of the numerical simulations and analytical derivations are presented in \jbfern{SM Sec.~IV}.

\subsection{Charging Dynamics: Numerical Results}
\label{sec:Charging3D}

$\\$\textbf{Ultrafast spreading dynamics:} To begin, Fig.~\ref{fig5_contour_plots} shows temporal snapshots of the charge density $\bar{\rho}$, salt concentration $\bar{C}$, and potential $\bar{\phi}$ fields near the patch during charging (also, see Movies \href{https://youtu.be/3kZTz5cyNl4}{S2}, \href{https://youtu.be/AVHCSiPRsUQ}{S3} and \href{https://youtu.be/u4_8NKB-Hzw}{S4}). These dynamics differ from the 1D case in two ways. First, up to a maximum pumping rate ${0 \leq \bar{j}^{\rm o}_0 < \bar{j}^{\rm o}_{\rm max}(\bar{R}^{\rm P})}$, the concentrations at all spatial locations are always non-negative for any $\bar{\tau}^{\rm o}$; the limit on the current $\bar{j}^{\rm o}_{\rm max}$ will be discussed later. Therefore, pumping can be continued indefinitely, as opposed to the 1D case, where open times $\bar{\tau}^{\rm o}$ are always bounded. Second,  near the pump, both $\bar{\rho}$ and $\bar{\phi}$ reach a local steady profile within a few Debye times, contrasting with the 1D case where both quantities continuously increase during charging  (see Fig.~\ref{fig3_uniform_pumping}). Note, however, that the 3D dynamics remain globally unsteady, as we shall now see.

\begin{figure}[t]
    \centering
    \includegraphics[width=\linewidth]{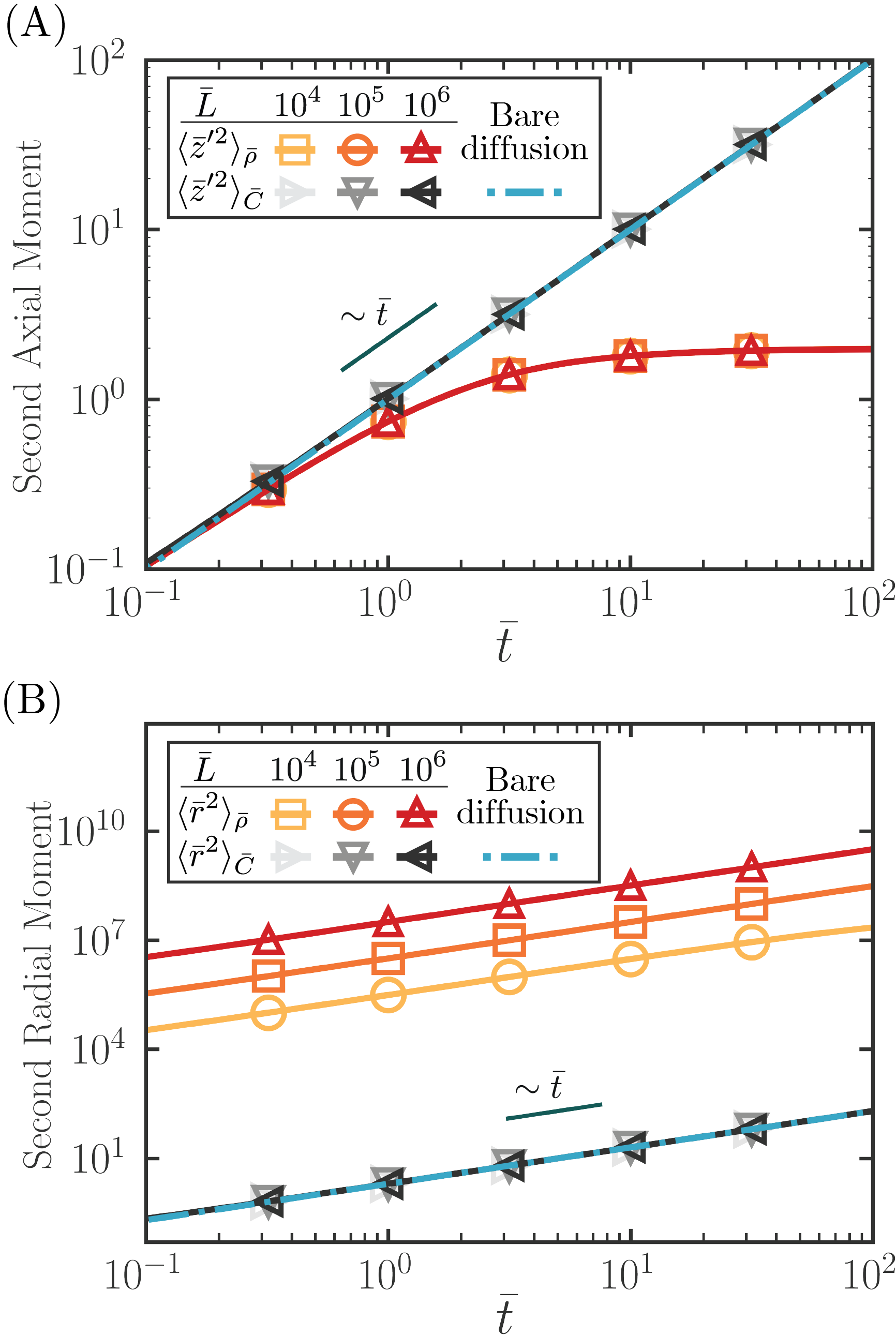}
    \caption{Time evolution of the axial and radial second moments of charge density and salt concentration distributions during continuous local pumping. Values from the FEM simulations (solid lines) are compared with bare diffusion of an uncharged solute (dash-dotted lines): ${\langle \bar{x}^2 \rangle = \gamma \bar{t}}$, where ${\gamma=1}$ and ${\gamma=2}$ for axial and radial diffusion, respectively (dashed lines). (A) The second axial moment for salt concentration evolves like bare diffusion. However, the axial moment of charge density saturates to $\sim1$, indicating that the pumped charges localize within a diffuse charge layer of thickness ${\sim \lambda_{\rm D}}$ near the membrane surface.
    (B) The second radial moments increase linearly with time. While the salt concentration evolves like bare diffusion in the radial direction, the charge density spreads orders of magnitude faster than the bare diffusion radially and exhibits dependence on the system size $\bar{L}$, i.e. ${\langle \bar{r}^2\rangle_{\bar{\rho}}=2\bar{D}_{\rm eff} (\bar{L}) \bar{t}}$.
    The finite size effect implies the long-range nature of in-plane charge reorganization.
    FEM simulation parameters as in Fig.~\ref{fig5_contour_plots}.
    A theoretical analysis of these observations is presented in Sec.~\ref{sec:second-moments}.
    }
\label{fig6_msd}
\end{figure}

To assess the spreading dynamics of $\bar{\rho}$ and $\bar{C}$, we consider the time-evolution of their second moments defined as
\begin{align}
     {\langle \bar{z}'^2 \rangle}_{\bar{\rm O}}(\bar{t}) &\eq \frac{\int_{\bar{\Omega}^{\rm II}} \bar{z}'^2\bar{\rm O} \, d\bar{\Omega}}{\int_{\bar{\Omega}^{\rm II}} \bar{\rm O} \, d\bar{\Omega}} \ , \\
     {\langle \bar{r}^2 \rangle}_{\bar{\rm O}}(\bar{t}) &\eq \frac{\int_{\bar{\Omega}^{\rm II}} \bar{r}^2\bar{\rm O} \, d\bar{\Omega}}{\int_{\bar{\Omega}^{\rm II}} \bar{\rm O} \, d\bar{\Omega}} \ ,     
    \label{eq:2nd_moment_defs}
\end{align}
where the observable $\bar{\rm O} \in \{\bar{\rho}, \bar{C}\}$ and ${\bar{z}'=\bar{z} -\bar{\delta}^{\rm M}/2}$ is the axial distance from the membrane surface. The integrals are evaluated in domain II, but one obtains the same values in domain I due to the symmetry of the pumping process. 

\begin{figure*}[t]
    \centering
    \includegraphics[width=\textwidth]{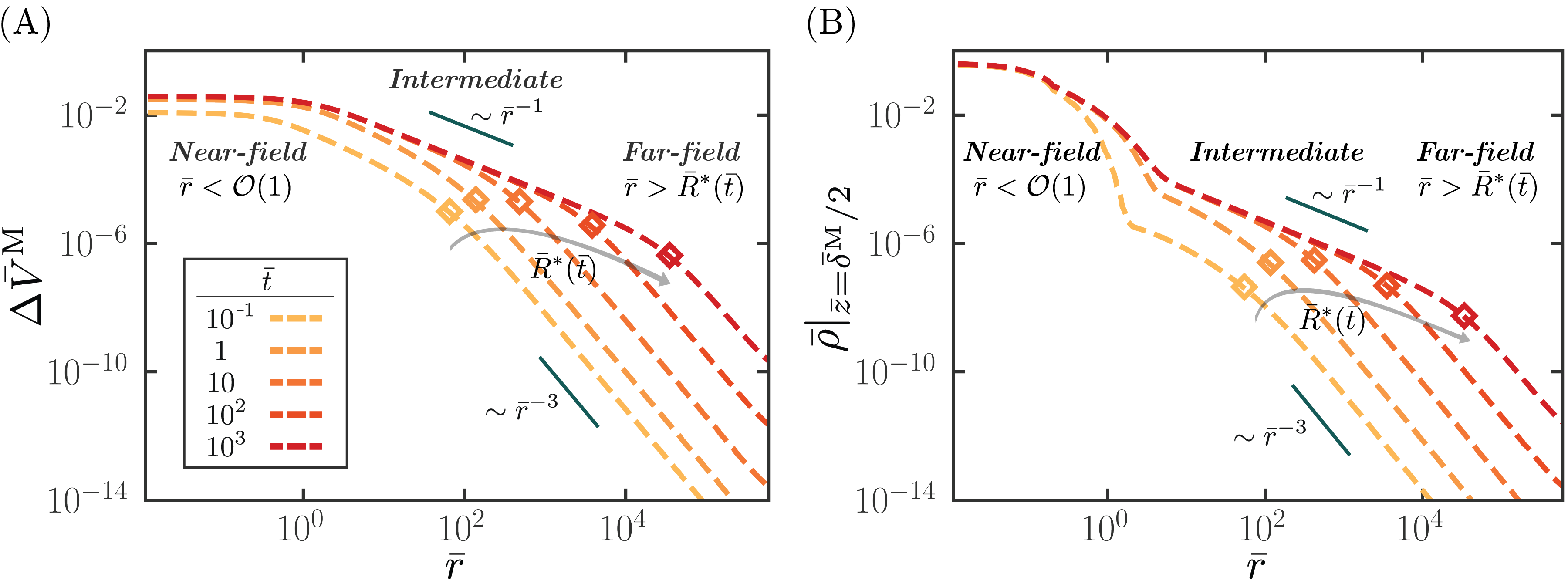}
    \caption{Time evolution of the radial profiles of (A) the transmembrane electric potential and (B) the charge density at the membrane surface (${\bar{z} = \bar{\delta}^{\rm M}/2}$) during localized pumping obtained by FEM. Indicated in each plot are the three regions with distinct scaling behaviors: near-field (${\bar{r}\lesssim 1}$), intermediate (${1\lesssim \bar{r}\lesssim \bar{R}^*)}$ with ${\bar{r}^{-1}}$ scaling, and far-field (${\bar{r}\gtrsim \bar{R}^*)}$ with {${\bar{r}^{-3}}$ scaling}. Markers indicate the transition location between the intermediate and far-field region ${\bar{R}^*\!\left(\bar{t}\right)}$. As pumping sustains, the intermediate region expands, and $\bar{R}^*(\bar{t})$ increases with time. Parameters used for the numerical simulations are as in Fig.~\ref{fig5_contour_plots}. }
\label{fig7_numerical_results}
\end{figure*}

Figures~\ref{fig6_msd}(A) and (B) show the axial and radial moments for concentration and charge density fields for various system sizes. For bare diffusion, we expect $ {\langle \bar{r}^2\rangle_{\bar{\rm O}} = 2\bar{t}}$ and ${\langle \bar{z}'^2 \rangle_{\bar{\rm O}} = \bar{t}}$. Figure ~\ref{fig6_msd} shows that while this is the case for $\bar{C}$, the axial and radial moments of $\bar{\rho}$ deviate significantly from the expectation of simple diffusion.
The axial moment $\langle \bar{z}'^2\rangle_{\bar{\rho}}$ saturates to a value of 2, indicating that a steady state is reached in the axial direction similar to the 1D case. In contrast, ${\langle \bar{r}^2\rangle_{\bar{\rho}} \approx 2\bar{D}_{\rm eff}(\bar{L}) \bar{t}}$, with an effective diffusivity ${\bar{D}_{\rm eff}(\bar{L}) \gg 1}$, for all sizes of the membrane $\bar{L}$ used in numerical simulations.
Not only is this effective diffusivity $\bar{D}_{\rm eff}(\bar{L})$ much larger than the self-diffusion coefficient, but also increases with system size $\bar{L}$ and would diverge for a truly infinite system.
These finite-size effects indicate the long-range nature of ionic reorganization in the radial direction. 
Thus, the in-plane charge reorganization due to pumping cannot be purely diffusive and raises a question as to the ultrafast nature of the spreading process.

$\\$\textbf{Three distinct regions of in-plane dynamics:}
\label{sec:three_regimes}
To further understand the nature of the ionic reorganization, we examine the time evolution of the transmembrane potential $\Delta \bar{V}^{\rm M}$ and the charge density at the membrane surface $\left. \bar{\rho} \right|_{{\bar{z}=\bar{\delta}^{\rm M}}/2}$.
As seen in Fig.~\ref{fig7_numerical_results}, the variation along $\bar{r}$ can be divided into three distinct regions: \emph{near-field}, \emph{far-field}, and \emph{intermediate}. The near-field region is defined by a hemispherical ``dome" of charge density $\bar{\rho}$, and a uniform $\Delta \bar{V}^{\rm M}$ for ${\bar{r}\lesssim 1}$. In the far-field region with ${\bar{r}\gtrsim \bar{R}^{*}(\bar{t})}$, both $\Delta \bar{V}^{\rm M}$ and $\bar{\rho}$ scale as $1/\bar{r}^{3}$. 
The far-field scaling of $\Delta \bar{V}^{\rm M}$ indicates that the electric potential field is similar to that induced by an electric dipole. 
By contrast, in the intermediate region, both $\Delta \bar{V}^{\rm M}$ and $\left. \bar{\rho} \right|_{{\bar{z}=\bar{\delta}^{\rm M}}/2}$ decay as $1/\bar{r}$, indicating that the electric potential field is similar to that induced by an electric monopole. Here, ${\bar{R}^{*}(\bar{t})}$ indicates the length scale for the crossover between intermediate and far-field regimes.
As is evident in Fig.~\ref{fig7_numerical_results}, the near-field and intermediate profiles reach a steady-state after ${\bar{t}\gtrsim 10}$, i.e. an order of magnitude of the Debye timescale, and all of the unsteady behavior is confined to the far-field. 
Moreover, as ions are actively pumped, the monopolar region expands radially at a steady rate $\dot{\bar{R}}^*$, which yields a speed of ${\sim40 \, {\rm m/s}}$ in physiological settings.

Taken together, the scaling behaviors from numerical simulations suggest active pumping induces long-ranged monopolar and dipolar electric fields, which in turn drive in-plane charge reorganization. To understand the underlying mechanisms behind the origin of the three distinct regimes, the nature of their steady or unsteady behaviors, and the resulting ultrafast spreading dynamics, we now turn to theoretical analysis. Given that a steady state is reached in a few Debye times in the near-field and intermediate regions, we proceed to study the dynamics by splitting the analysis into early ($\bar{t}<1$) and late time ($\bar{t}>1$) behaviors.

\subsection{Initial Response and Early Time Dynamics: Theoretical Analysis}
\label{sec:electrostatics}

\begin{figure}
\centering
\includegraphics[width=\linewidth]{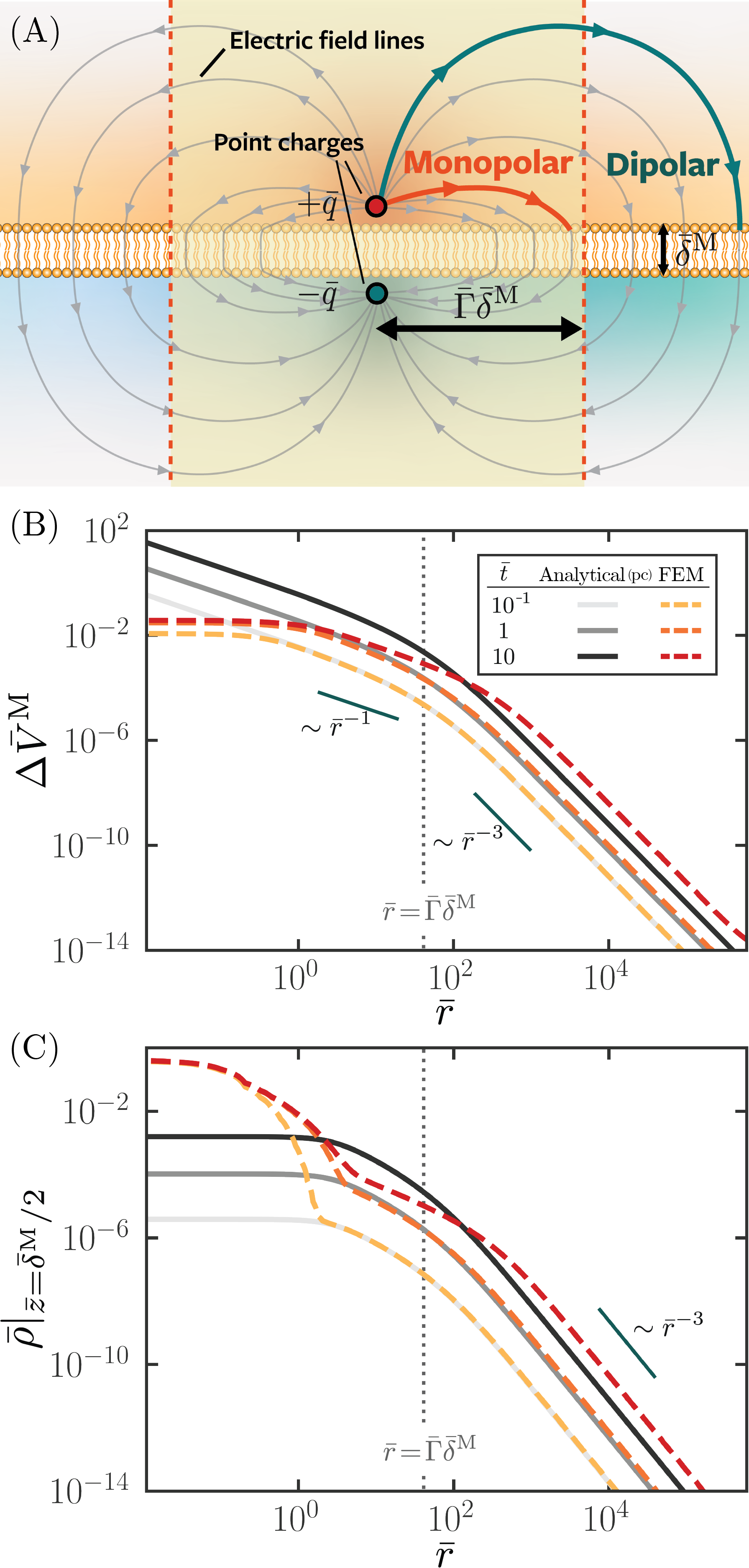}
    \caption{ (A) Schematic description of the intermediate and far-field regimes in short times. 
    Also shown are electric field lines induced by the point charges  in these regions.
    (B) Transmembrane electric potential and (C) charge density at the membrane. Analytical solutions (solid lines) obtained by the point charge approximation are compared with FEM simulations (dashed lines) at three time points. Physiological values are considered for the calculations: ${\bar{\Gamma}\!=\!20}$, ${\bar{\delta}^{\rm M}\!=\!4}$, ${\bar{j}^{\rm o}_0\!=\!4}$, and ${\bar{R}^{\rm P}\!=\!0.1}$.
    The comparisons confirm that the point charge approximation yields accurate solutions for short times ($\bar{t}\lesssim \mathcal{O}(1)$) in intermediate and far-field regions ($\bar{r}\gtrsim{O}(1)$).}
\label{fig8_electrostatics}
\end{figure}

$\\$\textbf{Point charge approximation:} During short times ${\bar{t}\lesssim 1\ (\approx 1\,{\rm ns})}$, there is limited diffuse charge reorganization throughout the system. At these early times, it may be reasonable to assume any charges that are pumped or removed remain concentrated around the transporter. This means that there are spatially fixed but equal and opposite point charges located on either side of the membrane (Fig.~\ref{fig8_electrostatics}(A)) of the form
\begin{equation}
    \bar{\rho}^{\rm pc}(\bar{r},\bar{z},\bar{t}) \eq  \frac{\bar{\delta}(\bar{r})}{2 \pi \bar{r}} \left[\bar{\delta}(\bar{z}-\bar{d}) - \bar{\delta}(\bar{z}+\bar{d})\right] \bar{q} (\bar{t})\ ,
    \label{eq:fixed_charge}
\end{equation}
where $\bar{d}$ ${(\approx \bar{\delta}^{\rm M}/2)}$ is the distance from the centerline to each point charge, $\bar{\delta}(\cdot)$ is the Dirac delta function and ${\bar{q}(\bar{t}) =\pi \left(\bar{R}^{\rm P}\right)^2 \bar{j}^{\rm o}_0\bar{t}}$ is the total pumped charge as a function of time. 
With no substantial diffuse charge reorganization, it may also be reasonable to approximate the electrolyte solutions in both domains as perfect dielectrics--- thus, there is no electroneutrality breakdown anywhere except at the location of the point charges. 
These fixed charges should then create instantaneous electric fields throughout out the system. 

$\\$\textbf{Emergence of long-range fields:} Given the point charge approximation, the corresponding electrostatic potential ${\bar{\phi}^{\rm pc}(\bar{r},\bar{z},\bar{t})}$ can be obtained exactly by solving Poisson's equation using the method of Hankel transform \jbfern{(SM Sec.~IV.1.(a))}. The full solution in domain II, for a large dielectric mismatch (${\bar{\Gamma}\gg 1}$), reduces to the following asymptotic form using the method of images
\begin{align}
 {\bar{\phi}}^{\rm pc, \rm II}  \eq 
  \begin{dcases}
    \frac{2\bar{\Gamma}}{1+\bar{\Gamma}}\frac{\bar{q}}{4\pi\bar{r}'} & \bar{r}'\ll \bar{\Gamma}\bar{\delta}^{\rm M} \ ,\\
    \frac{\bar{\Gamma}\bar{\delta}^{\rm M}\bar{q}(\bar{z}'+\frac{1}{2}\bar{\Gamma}\bar{\delta}^{\rm M})}{4\pi(\bar{r}^2+(\bar{z}'+\frac{1}{2}\bar{\Gamma}\bar{\delta}^{\rm M})^2)^{3/2}} & \bar{r}'\gg \bar{\Gamma}\bar{\delta}^{\rm M} \ ,\label{eq:elec-pot-monopole-dipole}
  \end{dcases}
\end{align} 
where ${\bar{r}'\equiv\sqrt{\bar{r}^2+\bar{z}'^2}}$ is the distance from the point source; see \jbfern{Fig.~S5} for the magnitudes and locations of the image charges.

The solution in Eq.~\eqref{eq:elec-pot-monopole-dipole} shows two distinct scalings of the electrical potential, with ${\bar{\Gamma}\bar{\delta}^{\rm M}}$ as the length scale for the transition location (Fig.~\ref{fig8_electrostatics}(A)).
In the region ${\bar{r}' \ll \bar{\Gamma}\bar{\delta}^{\rm M}}$,  the electric potential is that of a charge monopole of magnitude ${2\bar{q} \bar{\Gamma}/(1+\bar{\Gamma})\ (\approx 2\bar{q} \mbox{ for } \bar{\Gamma}\gg 1)}$. The factor of 2 arises from an image charge of the same sign located at the mirror position across the membrane-domain II interface (see \jbfern{Fig.~S5}). The effective potential is therefore a superposition of the pumped charge and its immediate image charge, yielding the effective $1/\bar{r}$ scaling along the plane of the membrane. 
On these length scales, one may equivalently say that the membrane, with its low dielectric permittivity, effectively screens the negative charge on the other side of the membrane in domain I.

On the other hand, in the region 
${\bar{r}' \gg\bar{\Gamma}\bar{\delta}^{\rm M}}$, Eq.~\eqref{eq:elec-pot-monopole-dipole} shows that the potential is that of a charge dipole with an enhanced dipole moment of 
${\bar{\Gamma}\bar{\delta}^{\rm M}\bar{q}}$ rather than ${\bar{\delta}^{\rm M}\bar{q}}$. This is because, for distances far away from the pump, a multipole expansion shows that the set of image charges may be represented as an effective dipole centered at ${\bar{z}_0=-(\bar{\Gamma}-1)\bar{\delta}^{\rm M}/2}$ along the axial direction \jbfern{(SM Sec.~IV.1.(c))}.  This dipole yields $1/\bar{r}^3$ scaling in the limit of large $\bar{r}$ along the membrane surface, and the presence of the low dielectric lipid membrane amplifies the potential by $\bar{\Gamma}^2$.

Altogether, the point charge approximation at these early times already provides an intuition for the emergence of the long-range monopolar and dipolar fields as observed in Fig.~\ref{fig7_numerical_results}. In particular, the presence of the low dielectric lipid membrane is crucial to the emergence of the monopolar regime---as can be seen from Eq.~\eqref{eq:elec-pot-monopole-dipole}, when $\bar{\Gamma}\to 1$ the transition length scale $\bar{\Gamma}\bar{\delta}^{\rm M}$ reduces to the membrane thickness and the monopolar regime diminishes.

$\\$\textbf{Intermediate and far-field dynamics at early times:}
The point charge approximation can be further used to examine the early-time dynamics observed in Fig.~\ref{fig7_numerical_results}.
To this end, the transmembrane potential can be easily estimated as ${\Delta \bar{V}^{\rm M}=\left.\bar{\phi}^{\rm pc, II} \right|_{S^{\rm II}} -  \left. \bar{\phi}^{\rm pc,I} \right|_{S^{\rm I}} }$. 
Figure~\ref{fig8_electrostatics}(B) shows that this simplified picture accurately predicts the early time behavior of ${\Delta \bar{V}^{\rm M}}$ observed in FEM simulations.  For large dielectric mismatch ${\bar{\Gamma}\gg1}$, Eq.~\eqref{eq:elec-pot-monopole-dipole} yields
\begin{align}
 \Delta\bar{V}^{\rm M}  \eq
  \begin{dcases}
    \frac{\bar{q}(t)}{\pi\bar{r}} & \bar{r}\ll \bar{\Gamma}\bar{\delta}^{\rm M} \ ,\\
    \frac{\bar{q}(t)(\bar{\Gamma}\bar{\delta}^{\rm M})^2}{4\pi\bar{r}^3} & \bar{r}\gg \bar{\Gamma}\bar{\delta}^{\rm M} \ ,\label{eq:trans-memb-electrostatics-early}
  \end{dcases}
\end{align} 
showing the monopolar and dipolar scalings in the intermediate and far-field regions, respectively. As expected, the point charge approximation produces errors near the patch (i.e. ${\bar{r}\lesssim 1}$). Moreover, for times ${\bar{t} > 1}$, the solution falls in accuracy even for ${\bar{r} \gtrsim 1}$ due to ionic reorganization throughout the system, which we shall analyze in the following section.

At these early times, much of the charge reorganization in the bulk domains is driven by the instantaneous electric fields created by the spatially fixed pumped point charges (see Fig.~\ref{fig8_electrostatics}(A) for the field lines). 
The charges in the solution migrate along the electric field lines and accumulate near the membrane-fluid interface. 
Given the long-range nature of the electric fields, scaling arguments suggest that charge gradients occur predominantly along the axial direction. In this case, the net charge accumulation should be driven by the electric fields normal to the membrane, leading to the emergence of diffuse layers.  This feature motivates a quasi-1D analysis to approximate the diffuse charge density $\bar{\rho}^{\rm II}$ \jbfern{(SM Sec.~IV.1.(d))}.

Figure~\ref{fig8_electrostatics}(C) shows that the quasi-1D analysis agrees well with the numerical results at early times. It also allows us to estimate the approximate charge density along the membrane surface as
\begin{equation}
    \bar{\rho}^{\rm II} |_{\bar{z}=\bar{\delta}^{\rm M}/2} \eq -\frac{4}{3}\sqrt{\frac{\bar{t}}{\pi}}
    \bar{E}_z^{\rm pc, II}|_{\bar{z}=\bar{\delta}^{\rm M}/2} + \mathcal{O}(\bar{t}^{5/2}) \ . \label{eq:chg_den_early}
\end{equation} 
The axial electric field, ${\bar{E}_z^{\rm pc, II}=-\partial \bar{\phi}^{\rm pc, II}/\partial \bar{z}}$, grows linearly with $\bar{t}$ with the deposited charge $\bar{q}$, and therefore, the charge density at the interface grows superlinearly, scaling as $\bar{t}^{3/2}$ at early times. In the limit of large $\bar{\Gamma}$, the electric field $\bar{E}_z^{\rm pc}$ at the surface can be shown to be \jbfern{(SM Sec.~IV.1.(d))}
\begin{equation}
    \bar{E}_z^{\rm pc, II}
    \eq \begin{dcases}
        -\frac{1}{4\pi}\frac{4\bar{\Gamma}}{\bar{\Gamma}^2-1}\frac{\bar{q}(\bar{t})}{(\bar{\delta}^{\rm M})^2}\frac{\pi^2}{6} & \bar{r}\ll \bar{\delta}^{\rm M} \ , \\
        -\frac{1}{4\pi}\frac{\bar{\Gamma}\bar{\delta}^{\rm M}\bar{q}(\bar{t})}{\bar{r}^3} & \bar{r}\gg \bar{\Gamma}\bar{\delta}^{\rm M} \ .
    \end{dcases} \label{eq:field-surface}
\end{equation}
Equation~\eqref{eq:field-surface} shows that, under the point charge approximation, the charge density decays as $1/\bar{r}^3$ far from the channel, consistent with the numerical results in Fig.~\ref{fig8_electrostatics}(C), but approaches a constant and fails in accuracy near the channel.

\begin{figure*}[t]
\centering\includegraphics[width=\textwidth]{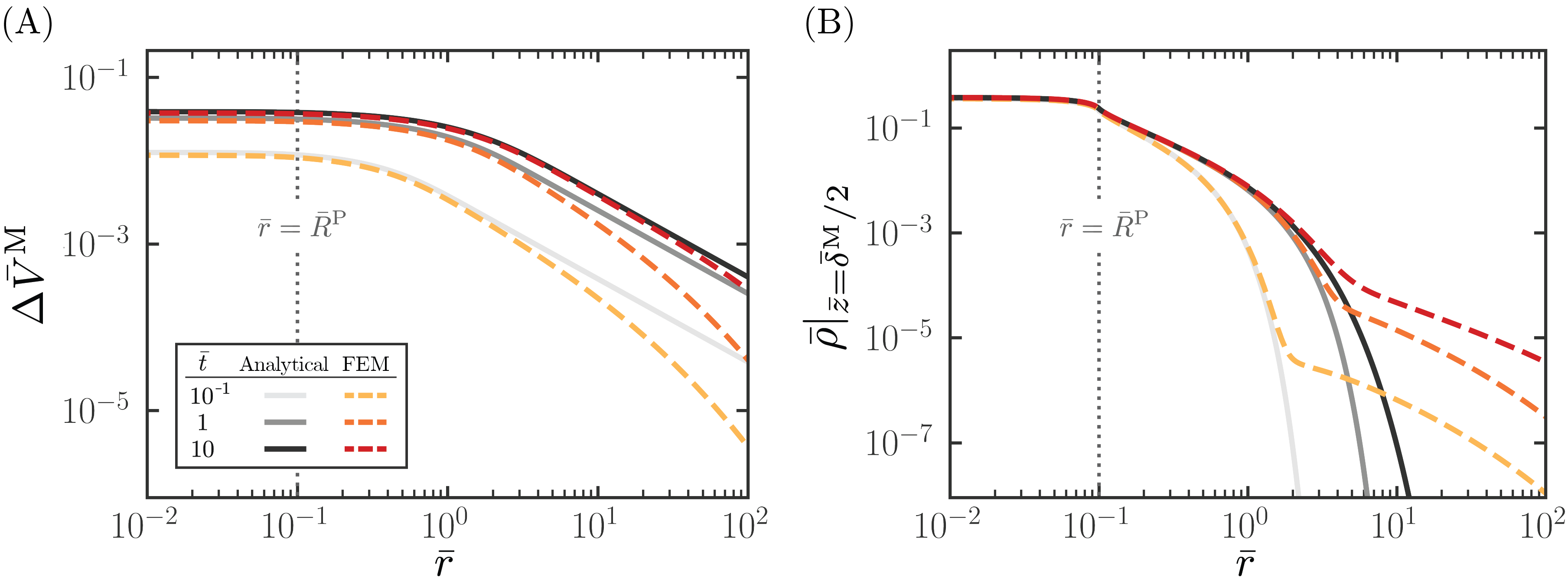}
    \caption{Near-field dynamics of (A) the transmembrane potential and (B) the charge density during continuous pumping of cations through a transporter of radius ${\bar{R}^{\rm P} = 0.1}$. Analytical solutions (solid lines) are compared with numerical solutions (dashed lines) obtained by FEM. The comparisons show that the analytical solutions accurately capture the near-field (${\bar{r}\lesssim 1}$) dynamics. In the near-field region, the transmembrane potential is flat while the charge density decays exponentially from the transporter. The near-field solutions demonstrate spherical symmetry about the center of the patch and reach steady states within $\mathcal{O}(1)$ Debye times. Physiological values are considered for all of the other parameters: ${\bar{\Gamma}\!=\!20}$, ${\bar{\delta}^{\rm M}\!=\!4}$, and ${\bar{j}^{\rm o}_0\!=\!4}$.}
\label{fig9_near_field}
\end{figure*}

While explaining the emergence of long-range fields, Figs.~\ref{fig8_electrostatics}(B) and (C) also make it clear that the point charge approximation fails to capture the near-field dynamics ($\bar{r}\lesssim 1$) at all times and the intermediate and far-field dynamics for $\bar{t}\gtrsim 1$. This is because this approximation ignores the ionic reorganization in the electrolyte solutions and its reciprocal influence on the electric potential, which we now address. 

\subsection{Late Time Dynamics: Theoretical Analysis}
\label{sec:late-time-dynamics}

$\\$\textbf{Near-field dynamics:} Despite its invalidity near the pump, the point charge picture still provides insight for obtaining the near-field solution. For ${\bar{r} \lesssim 1}$, the electrostatics in the near-field is dominated by the pumped charge $\bar{q}$ and its immediate image of charge ${\bar{q}{(\bar{\Gamma}-1)/(\bar{\Gamma}+1)} \approx \bar{q}}$ for ${\bar{\Gamma} \gg 1}$ \jbfern{(Fig.~S5)}. This image charge is located inside the membrane at an equal distance from the membrane surface as the pumped charge. As a consequence, the normal components of the electric fields from both charges cancel each other at the membrane-fluid interface near the pump. In this case, the boundary condition Eq.~\eqref{eq:nondim_bc_DConti} may be approximated in the near-field as
\begin{equation}
\bar{\bnabla}\,\bar{\phi}^\alpha\cdot\bbe_{z}\ \approx\ 0 \ . \label{eq:near_field_dielectric_BC}
\end{equation}
This also simplifies the flux boundary conditions Eqs.~\eqref{eq:nondim_bc_imposedFlux1} and~\eqref{eq:nondim_bc_imposedFlux2} at the membrane-fluid interface to
\begin{equation}
\label{eq:near_field_noflux_BC}
    -\bnabla \bar{\rho}^{\alpha} \cdot \bbe_{z} \eq
    -\bnabla \bar{C}^{\alpha} \cdot \bbe_{z} \eq \bar{j}^{\rm o}(\bar{r}, \bar{t}) \ .
\end{equation}

The new boundary conditions in Eqs.~\eqref{eq:near_field_dielectric_BC} and~\eqref{eq:near_field_noflux_BC} decouple the dynamics of the charge densities of the two domains $\bar{\rho}^{\rm I}$ and $\bar{\rho}^{\rm II}$ in the near-field, with the membrane serving as a perfect electrostatic screen. One can then solve for the charge density and electric potential in each domain independently.
To this end, we first consider a Green's function solution for a point source of cations at the pump, i.e. ${\bar{j}^{\rm o}=\bar{I} [\delta\left(\bar{r}\right) / (2\pi\bar{r})]}$ with ionic current ${\bar{I}=\int_{\bar{S}^{\rm II}} \bar{j}^{\rm o}\ d \bar{a}=\bar{j}_0^{\rm o}\pi (\bar{R}^{\rm P})^2}$. The ensuing transient analytical solutions for charge density and potential are cumbersome and are therefore provided in \jbfern{SM Sec.~IV.2.(b)}. 
Since ions are pumped through a patch of finite radius $\bar{R}^{\rm P}$, the charge density and potential are obtained by integration of the transient point source solutions.  As seen in Figs.~\ref{fig9_near_field}{(A) and (B)}, the resulting analytical solutions for the near-field are in good agreement with the nonlinear FEM solutions for all times.

From the analytical solutions, the near-field charge density relaxes to the following steady-state profile for $\bar{r}' \gtrsim \bar{R}^{\rm P}$:
\begin{equation}
    \bar{\rho}^{\rm II}_{\rm SS}
    \eq
     \frac{\bar{I}e^{-\bar{r}'}} {2\pi \bar{r}'} \ ,
 \label{eq:rho_nearfield}
\end{equation}
where $\bar{r}'$ is the distance from the center of the patch.
Equation~\eqref{eq:rho_nearfield} shows that during pumping, the flux of cations forms a hemispherical charge dome of length scale $\lambda_{\rm D}$, with a saturation value of  
${\bar{\rho}^{\rm II}|_{\bar{r}'=0}= \int_0^{\bar{R}^{\rm P}} \left(\bar{j}_0^{\rm o}e^{-\bar{r}'}/(2\pi \bar{r}') \right) \ 2\pi \bar{r}'\,d\bar{r}' \approx \bar{j}_0^{\rm o}\bar{R}^{\rm P}}$. 
Similarly, the potential relaxes to
\begin{equation}
    \bar{\phi}^{\rm II}_{\rm SS}
    \eq
    \frac{\bar{I}\left(1-e^{-\bar{r}'}\right)} {2\pi \bar{r}'}
    \ ,
    \label{eq:phi_nearfield}
\end{equation}
with a limiting value of $\bar{I}/2\pi$ at the center of the patch, consistent with the numerical results.
Note that the total charge inside the dome saturates to $\bar{I}$ within the Debye time (Fig.~\ref{fig9_near_field}). This shows that pumping of cations induces two steady-state hemispherical domes of size ${\sim \lambda_{\rm D}}$ containing equal and opposite charges of magnitude $\bar{I}$ on either side of the membrane.

$\\$\textbf{Bounds on current density and transmembrane potential at the pump:} The near-field solutions also reveal the maximum possible values for the current density $\bar{j}^{\rm o}_{\rm max}$ and the transmembrane potential at the pump. 
Ionic concentrations everywhere remain non-negative for all times as long as the imposed flux ${|\bar{j}^{\rm o}|<\bar{j}^{\rm o}_{\rm max} = (\bar{R}^{\rm P}+1-e^{-\bar{R}^{\rm P}})^{-1}\approx 1/(2\bar{R}^{\rm P})}$ for small $\bar{R}^{\rm P}$ \jbfern{(SM Sec.~IV.2.(c))}. 
The maximum allowable current density can be estimated as a balance between the imposed flux through the pore and the radial diffusive flux through a cylinder of radius $R^{\rm P}$ and height $\lambda_{\rm D}$ of the diffuse layer. This implies ${j_{\rm max}^{\rm o} \, \pi (R^{\rm P})^2 \sim (D C^0 / \lambda_{\rm D}) \, 2\pi R^{\rm P}\lambda_{\rm D}}$, yielding ${\bar{j}_{\rm max}^{\rm o}\sim 1/\bar{R}^{\rm P}}$ as a mass transfer limit for ionic currents across membranes. For physiological conditions, where the pore sizes typical of pumps and channels are small with ${\bar{R}^{\rm P} \approx 0.2 \  (2 \, \mathring{A})}$, the maximum allowable current is ${\bar{I}_{\rm max}\approx 10 \, \rm pA}$.
This mass transfer limited bound is close to typical currents measured through single channel recordings (${1-10\, \rm{pA}}$) \cite{hille1992,neher1976single,neher1992ion}.

When operating at the maximum current density $\bar{j}^{\rm o}_{\rm max}$, the transmembrane potential at the location of the pump quickly reaches its steady value
\begin{equation}
    \Delta\bar{V}^{\rm M}_{\rm max} \eq 2\frac{\bar{R}^{\rm P}-1+e^{-\bar{R}^{\rm P}}}{\bar{R}^{\rm P}+1-e^{-\bar{R}^{\rm P}}} \ ,
    \label{eq:maxV}
\end{equation}
which has an upper bound of $2$. Thus, the maximal $\Delta V^{\rm M}$ achievable by a localized single pump is ${2 k_B T / \rm{e} \approx 50 {\rm mV}}$. 
However, for small pore sizes of transporters, ${\Delta\bar{V}^{\rm M}_{\rm max}\approx \bar{R}^{\rm P}/2}$. For ${\bar{R}^{\rm P} \approx 0.2}$, we may obtain maximum achievable potential differences of ${\sim2.5 \, {\rm mV}}$, much smaller than typical gating potentials of ion channels \cite{hodgkin1952measurement,hille1992}.

\begin{figure}[t]
    \centering
    \includegraphics[width=\linewidth]{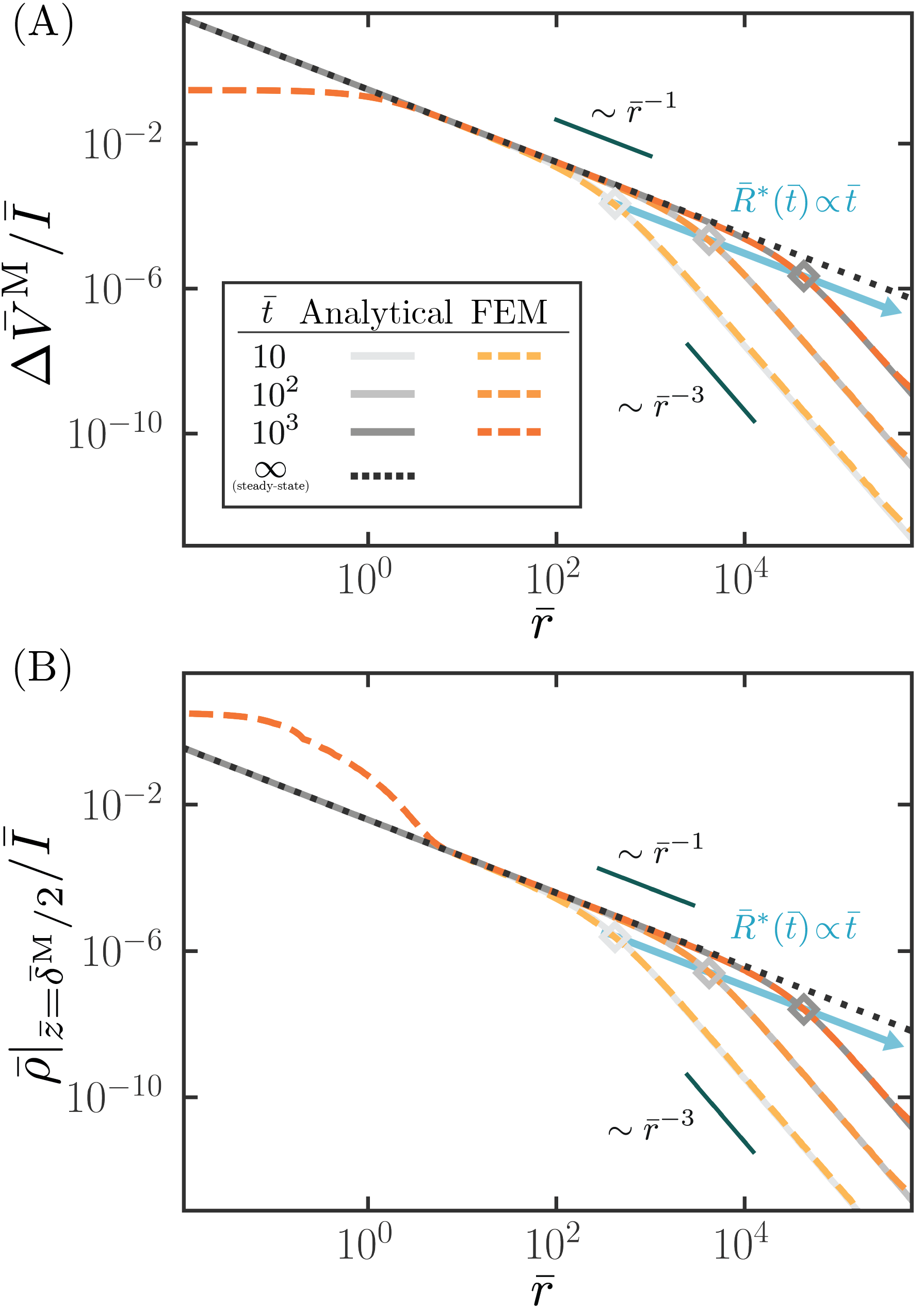}
    \caption{Analytical solutions capture the late-time dynamics in the intermediate and far-field regions, exhibiting the monopolar and dipolar character, respectively. Panel (A) shows the transmembrane potential scaled by the ionic current, with agreement for $\bar{t},\bar{r}\gtrsim 1$ between the numerical and analytical solutions for $\bar{r}\gtrsim 1$. The dotted line represents the analytical steady-state solution. Panel (B) shows the same for the charge density at the surface. 
    }
\label{fig10_local_pumping}
\end{figure}

$\\$\textbf{Intermediate and far-field dynamics:}
We now discuss the late-time dynamics of intermediate and far-field regions, as observed in Fig.~\ref{fig7_numerical_results}. The linearized PNP equations can be solved exactly using Hankel and Laplace transforms in the radial and time domains, respectively (see \jbfern{SM Sec.~IV.3.(a)} for full analytical solutions). 
With continuous localized pumping, the charge density and transmembrane potential for ${\bar{r}\gg1}$ reach steady state profiles given by 
\begin{align}
\bar{\rho}_{\textrm{SS}}^{\rm II}(\bar{r}, \bar{z}) \eq &\frac{\bar{I}}{2\pi}\frac{e^{-\bar{z}'}}{\bar{v}\bar{r}} \ , \label{eq:rhoSS}\\
\Delta\bar{V}_{\textrm{SS}}^{\rm M}(\bar{r}) \eq &\frac{\bar{I}}{\pi\bar{r}}\left(1-\frac{1}{\bar{v}}\right) \ , \label{eq:phiSS}
\end{align}
where ${\bar{v}=1+\bar{\Gamma}\bar{\delta}^{\rm M}/2}$. The steady-state charge density decays exponentially with $\bar{z}'$, consistent with the formation of a diffuse charge layer of length scale $\lambda_{\rm D}$. 
The expressions for $\bar{\rho}_{\textrm{SS}}^{\rm II}$ and $\Delta\bar{V}_{\textrm{SS}}^{\rm M}$ recover the $1/\bar{r}$ monopolar scaling and agree well with the numerical solutions in the intermediate region, as shown in Fig.~\ref{fig10_local_pumping}. Moreover, for a large dielectric mismatch, ${1 / \bar{v}\ll 1}$, and the steady-state transmembrane potential in Eq.~\eqref{eq:phiSS} reduces to the monopolar potential of the point charge approximation in Eq.~\eqref{eq:trans-memb-electrostatics-early}. Here, the two steady-state hemispherical domes act as equal and opposite point charges of magnitude $\bar{I}$ with the membrane as an electrostatic screen. 

As for the transient far-field solutions, numerical results in Fig.~\ref{fig7_numerical_results} suggest that the transition location $\bar{R}^{*}(\bar{t})$ increases linearly with time, which is further confirmed by a universal collapse of the scaled potential ${\bar{t}\bar{\phi}}$ with ${\bar{r}/\bar{t}}$  in the intermediate and far-field regions \footnote{Although reminiscent of self-similarity, standard methods of identifying similarity solutions fail here as the near-field portion is not self-similar \jbfern{(Fig.~S6)}.} \jbfern{(Fig.~S6)}.
Motivated by this linear scaling ${\bar{R}^{*}\propto \bar{t}}$, we consider the limit ${\bar{r},\bar{t}\rightarrow \infty}$ with fixed ${\bar{r}/\bar{t}}$ and obtain the following asymptotic solutions
\begin{align}
    \bar{\rho}^{\rm II}(\bar{r}, \bar{z}, \bar{t}) \ &\approx\  \bar{\rho}_{\textrm{SS}}^{\rm II}(\bar{r}, \bar{z})\left[1-\left(1+\left(\frac{\bar{v}\bar{t}}{\bar{r}}\right)^2 \right)^{-\frac{1}{2}} \right] \ , \label{eq:rho_ff}\\
    \Delta\bar{V}^{\rm M}(\bar{r}, \bar{t})  \ &\approx\  \Delta\bar{V}_{\textrm{SS}}^{\rm II}(\bar{r})\left[1-\left(1+\left(\frac{\bar{v}\bar{t}}{\bar{r}}\right)^2 \right)^{-\frac{1}{2}} \right] \ . \label{eq:phi_ff}
\end{align}
Figure~\ref{fig10_local_pumping} confirms that these transient solutions accurately predict the dynamics of the full PNP equations in the intermediate and far-field regions at long times. For ${\bar{r}\ll \bar{v}\bar{t} }$, Eqs.~\eqref{eq:rho_ff} and~\eqref{eq:phi_ff} trivially recover the steady-state solutions in Eqs.~\eqref{eq:rhoSS} and~\eqref{eq:phiSS} which describe the observed ${1/\bar{r}}$ monopolar profile in the intermediate region. On the other hand, for ${\bar{r}\gg \bar{v}\bar{t}}$ we have 
\begin{align}
    \bar{\rho}^{\rm II}(\bar{r}, \bar{z}, \bar{t}) \ &\approx\ \frac{\bar{I}}{4\pi} \frac{\bar{v} \bar{t}^2}{\bar{r}^3} e^{-\bar{z}'}\ , \\
    \Delta\bar{V}^{\rm M}(\bar{r}, \bar{t}) \ &\approx \ \frac{\bar{I}}{2\pi}\left(1-\frac{1}{\bar{v}}\right) \frac{\bar{v}^2\bar{t}^2}{\bar{r}^3} \ ,
\end{align}
recovering the observed $1/\bar{r}^3$ dipolar scaling of the far-field.

$\\$\textbf{Expansion speed of monopolar region:} Equations~\eqref{eq:rho_ff} and~\eqref{eq:phi_ff} show the crossover length scale between the intermediate and far-field regions is ${\bar{R}^{*} (\bar{t})=\bar{v} \bar{t}}$. Its rate ${\dot{\bar{R}}^{*}=\bar{v}}$ represents the expansion speed of the monopolar region, which in dimensional units corresponds to a speed of
\begin{equation}
    v \eq \left(1+\frac{\bar{\Gamma}\bar{\delta}^{\rm M}}{2}\right)\frac{D}{\lambda_{\rm D}} = \left(1+\frac{1}{2} \frac{\epsilon}{\epsilon^{\rm M}}\frac{\delta^{\rm M}}{\lambda_{\rm D}} \right) \frac{D}{\lambda_{\rm D}} \ .
    \label{eq:crossover_v}
\end{equation}
For physiological conditions where ${\bar{\Gamma} \sim 20}$ and ${\bar{\delta}^{\rm M} \sim 4}$, we find ${v \sim 40\, {\rm m/s}}$. 
This indicates that for distances ${r \gg \lambda_{\rm D}}$, the monopolar region spreads at a rate that far exceeds bare diffusion. For length scales of $1\,\si{\micro m}$, bare diffusion requires a time scale of order ${\left(1\,\si{\micro m}\right)^2/D\sim 1\, \si{ms}}$ that is orders of magnitude slower than the monopolar spreading time, ${1\, \si{\micro m}/v\sim 25 \, \si{ns}}$.
Equation~\eqref{eq:crossover_v} also suggests that the rate of radial expansion of the monopolar region is enhanced by the dielectric mismatch and membrane thickness with the factor $\bar{\Gamma}\,\bar{\delta}^{\rm M}/2$.
 
\subsection{Second Moments and Ultrafast Dynamics}
\label{sec:second-moments}
Using results presented in the previous subsections, we now obtain analytical expressions that explain the scalings observed in the axial and radial second moments of $\bar{\rho}$ in Fig.~\ref{fig6_msd}; see \jbfern{SM Sec.~IV.4} for detailed derivations.

$\\$\textbf{Early times:} At early times, the development of the charge dome dominates the axial reorganization.
Using the full transient solution for the charge dome, we find ${\langle \bar{z}'^2\rangle_{\bar{\rho}}\approx\bar{t}}$, consistent with the bare diffusion result in Fig.~\ref{fig6_msd}(A).
In contrast, the reorganization of charge in the radial direction is primarily due to the development of the diffuse charge layer along the membrane-fluid surfaces.
For a large system size 
${\bar{L} \gg\bar{\Gamma}\bar{\delta}^{\rm M}}$, charge reorganization in the dipolar region forms the main contribution yielding
\begin{equation}
    \langle \bar{r}^2\rangle_{\bar{\rho}} \ \approx\  \frac{1}{4} \bar{L} \bar{\Gamma}\bar{\delta}^{\rm M}\bar{t} \ .\label{eq:radial_moment_early}
\end{equation}
with an effective diffusivity of $\bar{L} \bar{\Gamma}\bar{\delta}^{\rm M}/8$. This is consistent with the system size dependence noted in Fig.~\ref{fig6_msd}(B) and demonstrates the role of long-range effects in charge reorganization.

$\\$\textbf{Late times:} Since the charge dome saturates within the Debye time, the dynamics at late times are primarily from the charge reorganization in the intermediate and far-field regimes.
At these length and time scales, charges reside near membrane surfaces, forming diffuse charge layers of thickness $\lambda_{\rm D}$. Using the asymptotic solution Eq.~\eqref{eq:rho_ff}, we find ${\langle \bar{z}'^2\rangle_{\bar{\rho}}=2}$, which is consistent with the numerical results shown in Fig.~\ref{fig6_msd}(A). This indicates that the axial charge reorganization is confined to diffuse charge layers at long times.

As for the radial moment, the charge reorganization is entirely due to the transient far-field region where the charge density scales as $1/\bar{r}^3$. From the far-field solution Eq.~\eqref{eq:rho_ff} we find
\begin{equation}
    \langle  \bar{r}^2\rangle_{\bar{\rho}}\ \approx \  \frac{1}{2}\bar{L} \bar{v}  \bar{t} \ . \label{eq:radial_moment_late}
\end{equation}
Equation~\eqref{eq:radial_moment_late} is again consistent with Fig.~\ref{fig6_msd}(A) at late times, with a system size dependent effective diffusivity 
${\bar{D}_{\rm eff} = \bar{L}\bar{v}/4}$, which is proportional to the monopolar spreading speed. Moreover, for a large dielectric mismatch the effective diffusivity ${\bar{D}_{\rm eff} = \bar{L} \bar{v}/4 \approx \bar{L}\bar{\Gamma}\bar{\delta}^{\rm M} / 8}$ matches with early time result in Eq.~\eqref{eq:radial_moment_early}, showing consistent behaviors at all times.
Taken together, the scaling results suggest that long-ranged electric fields cause ultrafast charge rearrangements in the radial direction while being confined near the membrane surfaces.

\begin{figure}[t]
    \centering
    \includegraphics[width=\linewidth]{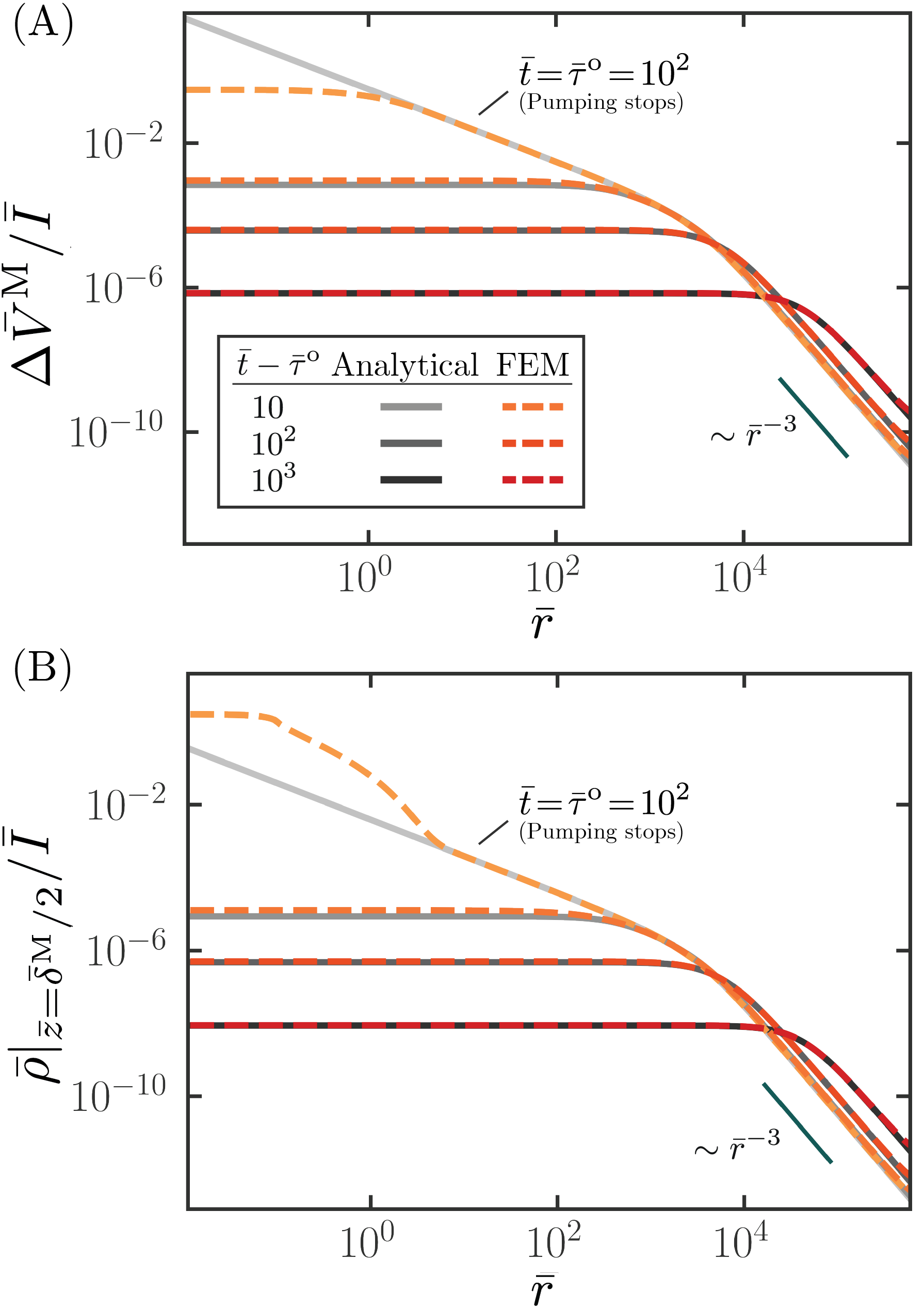}
    \caption{Relaxation dynamics after deactivation of cationic pumping at $\bar{t}\!=\!\bar{\tau}^{\rm o}\!=\!10^2$. Within a few Debye times, both (A) transmembrane potential and (B) charge density at the surface form plateaus in the intermediate region, which still expands with the speed $\bar{v}$. In the far field, the dipolar $1/\bar{r}^3$ scaling remains while it grows slower with time compared to when pumping is in action. The analytical solutions are obtained by the superposition principle, where we subtract a time-shifted copy of the long-time asymptotic solutions from itself. Due to the asymptotic nature of the analytical solution, its accuracy increases with time. Both the transmembrane potential and the charge density at the surface are scaled by the ionic current.
    }
\label{fig_local_pump_relax}
\end{figure}

\subsection{Relaxation Dynamics}\label{sec:relaxation-3D}
As a last feature of the ionic reorganization dynamics, we now discuss how the system relaxes upon deactivation of pumping.
Figure~\ref{fig_local_pump_relax} shows FEM results for the transmembrane potential and charge density once the pumping stops at ${\bar{t} = \bar{\tau}^{\rm o} = 10^2}$. Noticeably, within a few Debye times after deactivation, the charge domes and the intermediate region with the $1/\bar{r}$ (monopolar) scaling disappear. The profiles of both the charge density and electric potential in these regions exhibit plateaus, whose values decrease towards a state of uniformity across the entire system. The far-field charge and potential, however, maintain the dipolar ${1/\bar{r}^3}$ scaling, but grow more slowly with time than they do during pumping. Lastly, as the relaxation proceeds, the plateau region continues to expand with the same speed $\bar{v}$ even after deactivation.

We obtain analytical solutions for the relaxation dynamics by superposing time-shifted copies of the charging solutions in Eqs.~\eqref{eq:rho_ff} and~\eqref{eq:phi_ff} \jbfern{(SM Sec.~IV.5)}.
These analytical solutions agree well with the numerical results, as shown in Fig.~\ref{fig_local_pump_relax}. In the radially uniform plateau region (${1\ll \bar{r} \ll \bar{R}^{*}(\bar{t} - \bar{\tau}^{\rm o})}$), the charge density and the membrane potential are
\begin{align}
\bar{\rho}^{\rm II}(\bar{r}, \bar{z}, \bar{t})\ &\approx \ \frac{\bar{I}\bar{\tau}^{\rm o}}{2 \pi (\bar{v}\bar{t})^{2}}e^{-\bar{z}'}
 \ , \label{eq:rho_rlx_plt}\\
\Delta\bar{V}^{\rm M}(\bar{r}, \bar{t})\ &\approx \ \frac{\bar{I}\bar{\tau}^{\rm o}}{\pi (\bar{v}\bar{t})^{2}}(\bar{v} - 1)\ , \label{eq:phi_rlx_plt}
\end{align}
which decay quadratically in time. 
The corresponding far-field (${{\bar{r} \gg \bar{R}^{*}(\bar{t}})}$) solutions are given by
\begin{align}
    \bar{\rho}^{\rm II}(\bar{r}, \bar{z}, \bar{t})  \ &\approx \ \frac{\bar{I}\bar{\tau}^{\rm o}}{2\pi} \frac{\bar{v} \bar{t}}{\bar{r}^3} e^{-\bar{z}'}\ , \label{eq:rho_rlx_far}\\
    \Delta\bar{V}^{\rm M}(\bar{r}, \bar{t})  \ &\approx \ \frac{\bar{I}\bar{\tau}^{\rm o}}{\pi}
    \frac{\bar{t}}{\bar{r}^3} (\bar{v} - 1)\ , \label{eq:phi_rlx_far}
\end{align}
which show the dipolar scaling $1/\bar{r}^3$ and grow linearly in time, consistent with the numerical results.

Once specific ions move across a membrane, charge neutrality is broken in individual electrolyte solutions and never recovered. Due to the global charge neutrality of the whole system, the superposition of electric potentials generated by the two electrolyte solutions becomes dipole-like at a sufficiently far distance. Therefore, the dipolar character should remain at far distances whether pumping stops or not. The resulting electric fields then cause electromigration, which accumulates charges at the membrane interfaces with the charge density following the ${1/\bar{r}^3}$ scaling of the electric field normal to the interface. In contrast, the charge domes and the monopolar scalings in the intermediate region are a direct consequence of sustained pumping (Sec.~\ref{sec:late-time-dynamics}) and vanish upon deactivation.

\section{Conclusions and Implications}

We study ionic reorganization dynamics near biological membranes by considering the localized transmembrane transport of cations, typical of ion channels and pumps in biological systems. The localized transport is modeled as a coupled source and sink on either side of the membrane. The ensuing charge separation breaks local electroneutrality, leading to long-range electric fields that drive 3D ionic reorganization at the membrane-fluid interfaces. Spatiotemporal dynamics are analyzed using theory and large-scale FEM simulations under the PNP framework, explicitly accounting for the ion transport phenomena commonly neglected in classical cable theoretic models.

As the localized transporter operates, the charge rearrangements occur via the emergence of three distinct regions along the membrane-fluid interfaces: a steady-state charge dome near the source, a radially expanding intermediate monopolar region, and a transient far-field dipolar region. Electroneutrality breakdown is limited to a few Debye lengths on either side of the membrane, corresponding to the diffuse charge layers of equal and opposite net charge density.  Upon continuous pumping, the intermediate region, where electric fields decay more slowly than in the far-field, expands in the radial direction along the membrane fluid interface with a steady speed. We estimate the speed of expansion to be ${\sim 40 \, \si{m/s}}$ in physiological conditions, reminiscent of the typical speeds of action potential propagation in neuronal membranes \cite{hodgkin1952quantitative,hille1992, hodgkin1939action}. 

An important conclusion of our work is that an active transmembrane ionic flux results in long-ranged electric fields, as manifested in the monopolar region. Moreover, the size and speed of expansion of this region is enhanced by the dielectric mismatch and finite thickness of the lipid membrane. The electrostatic interactions mediated by these long-ranged fields may be important to a number of coupled biological processes driven by ion transport across membranes. However, biological systems are far more complex than the ones considered in this work. Part of the complexity arises from the multicomponent and asymmetric nature of the electrolyte solutions with varying valencies, diffusivities, and bulk concentrations of ionic species. Moreover, the membrane may itself carry surface charges in the form of charged lipid head groups. Other complexity arises from the intrinsic geometry of the membranes \cite{sahu2020geometry}, may it be their existence in the form of spherical vesicles or cylindrical shapes in neuronal dendrites and axons. Nevertheless, the numerical and theoretical framework presented here may be a starting point to address such complexity, and merits future development.

We end with a calculation that illustrates a potential role for long-range electrostatic effects in determining the length scales involved in neuronal excitation. During action potential generation, it is known that transmembrane potential must change by at least ${10\, \si{mV}}$ \cite{hodgkin1952measurement,hille1992}. However, the estimated maximum transmembrane potential achievable by a single source ${\Delta\bar{V}^{\rm M}_{\rm max} \approx 2.5\, \si{mV}}$ (Sec.~\ref{sec:late-time-dynamics}) suggests that a single transporter is insufficient and cooperation among several transporters may be required to achieve such biologically relevant potential differences. This motivates the question: for a specified density of such transporters, what is the minimum size of a membrane patch that should be excited to achieve ${\sim 10\, \si{mV}}$ potential difference over the entire patch?

The patch size can be readily estimated by using the analytical solutions presented in this work. Suppose the transporters are distributed in a circular patch of radius $R$ with a density $\chi$ (i.e. channels per area of the patch). Upon non-dimensionalization ${\bar{R}=R/\lambda_{\rm D}}$ and ${\bar{\chi}=\chi \lambda_D^2}$ and supposing that ${\bar{t}>\bar{R}/\bar{v}}$, we may assume that the system reaches a steady-state within the patch. Additionally, for ${\bar{R} \gg 1}$, the steady state corresponds to the intermediate monopolar region. Assuming that the excitation of the patch entails the transporters acting in concert, the net transmembrane potential difference at the center of the patch is a superposition of the potential differences due to individual transporters, given by Eq.~\eqref{eq:phiSS}. Integrating over the patch area yields ${\Delta\bar{V}^{\rm M} = 2\bar{\chi} \bar{R} \bar{I} (1 - 1/\bar{v})}$, with $\bar{I}$ being the current for a single channel. Dimensionalizing this result yields
\begin{equation}
    \Delta V^{\rm M} \eq \frac{k_{\rm B}T}{e^2 D C_{\rm 0}}  I \chi R \left(\frac{C_{\rm D}}{2C_{\rm M} + C_{\rm D}}\right) \ , \label{eq:excitability}
\end{equation}
showing the transmembrane potential scales linearly with the patch radius as a direct consequence of the monopolar electrostatic contributions.
Because ${C_{\rm D} \gg C_{\rm M}}$, the term in the parentheses is approximately 1. This gives, 
\begin{equation}
    \Delta V^{\rm M}  \approx \ 
    \left(I \chi \pi R^2 \right)
    \left(\frac{k_{\rm B}T}{\pi e^2 D C_{\rm 0} R} \right)
    \eq 
    \mathcal{I}_{\rm patch}  \mathcal{R}_{\rm patch} \ , \label{eq:excitability2}
\end{equation}
recovering a current-voltage relationship for the patch for an overall current $\mathcal{I}_{\rm patch} = I \chi \pi R^2$, and an effective resistance $\mathcal{R}_{\rm patch}$ that decreases with the patch size $R$.

Given the current ${I \sim 10 \, \si{pA}}$ from single channel recordings \cite{neher1976single} with channel densities of ${\chi \sim 10^3 \, \si{\micro m}^{-2}}$ \cite{rosenbluth1976intramembranous}, Eq.~\eqref{eq:excitability} shows that achieving a physiological ${\Delta V^{\rm M} \sim 25\, {\si{mV}}}$ requires a minimum patch of radius ${R\sim 1.6 \, \si{\micro m}}$. Predictions of the minimum length scales required for excitability and their dependence on the density of channels as in Eq.~\eqref{eq:excitability} may be experimentally tested.

\begin{acknowledgments}
J.B.F. acknowledges support from the U.S. Department of Energy, Office of Science, Office of Advanced Scientific Computing Research, Department of Energy Computational Science Graduate Fellowship under Award Number DE-SC0023112.
K.K.M is supported by Director, Office of Science, Office of Basic Energy Sciences, of the U.S. Department of Energy under contract No. DEAC02-05CH11231. 
K.S. acknowledges support from the Hellman Foundation, the McKnight Foundation, and the University of California, Berkeley. 
This research used resources of the National Energy Research Scientific Computing Center (NERSC), a U.S. Department of Energy Office of Science User Facility located at Lawrence Berkeley National Laboratory, using NERSC award BES-ERCAP0023682.
\end{acknowledgments}


\bibliography{references}

\begin{thebibliography}{52}%
\makeatletter
\providecommand \@ifxundefined [1]{%
 \@ifx{#1\undefined}
}%
\providecommand \@ifnum [1]{%
 \ifnum #1\expandafter \@firstoftwo
 \else \expandafter \@secondoftwo
 \fi
}%
\providecommand \@ifx [1]{%
 \ifx #1\expandafter \@firstoftwo
 \else \expandafter \@secondoftwo
 \fi
}%
\providecommand \natexlab [1]{#1}%
\providecommand \enquote  [1]{``#1''}%
\providecommand \bibnamefont  [1]{#1}%
\providecommand \bibfnamefont [1]{#1}%
\providecommand \citenamefont [1]{#1}%
\providecommand \href@noop [0]{\@secondoftwo}%
\providecommand \href [0]{\begingroup \@sanitize@url \@href}%
\providecommand \@href[1]{\@@startlink{#1}\@@href}%
\providecommand \@@href[1]{\endgroup#1\@@endlink}%
\providecommand \@sanitize@url [0]{\catcode `\\12\catcode `\$12\catcode `\&12\catcode `\#12\catcode `\^12\catcode `\_12\catcode `\%12\relax}%
\providecommand \@@startlink[1]{}%
\providecommand \@@endlink[0]{}%
\providecommand \url  [0]{\begingroup\@sanitize@url \@url }%
\providecommand \@url [1]{\endgroup\@href {#1}{\urlprefix }}%
\providecommand \urlprefix  [0]{URL }%
\providecommand \Eprint [0]{\href }%
\providecommand \doibase [0]{https://doi.org/}%
\providecommand \selectlanguage [0]{\@gobble}%
\providecommand \bibinfo  [0]{\@secondoftwo}%
\providecommand \bibfield  [0]{\@secondoftwo}%
\providecommand \translation [1]{[#1]}%
\providecommand \BibitemOpen [0]{}%
\providecommand \bibitemStop [0]{}%
\providecommand \bibitemNoStop [0]{.\EOS\space}%
\providecommand \EOS [0]{\spacefactor3000\relax}%
\providecommand \BibitemShut  [1]{\csname bibitem#1\endcsname}%
\let\auto@bib@innerbib\@empty
\bibitem [{\citenamefont {Lodish}(2000)}]{lodish2000molecular}%
  \BibitemOpen
  \bibfield  {author} {\bibinfo {author} {\bibfnamefont {H.}~\bibnamefont {Lodish}},\ }\href@noop {} {\emph {\bibinfo {title} {Molecular Cell Biology}}}\ (\bibinfo  {publisher} {W.H. Freeman},\ \bibinfo {year} {2000})\BibitemShut {NoStop}%
\bibitem [{\citenamefont {Hille}(1992)}]{hille1992}%
  \BibitemOpen
  \bibfield  {author} {\bibinfo {author} {\bibfnamefont {B.}~\bibnamefont {Hille}},\ }\href@noop {} {\emph {\bibinfo {title} {Ionic Channels of Excitable Membranes}}}\ (\bibinfo  {publisher} {Oxford University Press, Incorporated},\ \bibinfo {year} {1992})\BibitemShut {NoStop}%
\bibitem [{\citenamefont {Hodgkin}\ and\ \citenamefont {Katz}(1949)}]{hodgkin1949effect}%
  \BibitemOpen
  \bibfield  {author} {\bibinfo {author} {\bibfnamefont {A.~L.}\ \bibnamefont {Hodgkin}}\ and\ \bibinfo {author} {\bibfnamefont {B.}~\bibnamefont {Katz}},\ }\bibfield  {title} {\bibinfo {title} {The effect of sodium ions on the electrical activity of the giant axon of the squid},\ }\href@noop {} {\bibfield  {journal} {\bibinfo  {journal} {The Journal of Physiology}\ }\textbf {\bibinfo {volume} {108}},\ \bibinfo {pages} {37} (\bibinfo {year} {1949})}\BibitemShut {NoStop}%
\bibitem [{\citenamefont {Hodgkin}\ and\ \citenamefont {Huxley}(1952)}]{hodgkin1952quantitative}%
  \BibitemOpen
  \bibfield  {author} {\bibinfo {author} {\bibfnamefont {A.~L.}\ \bibnamefont {Hodgkin}}\ and\ \bibinfo {author} {\bibfnamefont {A.~F.}\ \bibnamefont {Huxley}},\ }\bibfield  {title} {\bibinfo {title} {A quantitative description of membrane current and its application to conduction and excitation in nerve},\ }\href@noop {} {\bibfield  {journal} {\bibinfo  {journal} {The Journal of Physiology}\ }\textbf {\bibinfo {volume} {117}},\ \bibinfo {pages} {500} (\bibinfo {year} {1952})}\BibitemShut {NoStop}%
\bibitem [{\citenamefont {Cole}\ and\ \citenamefont {Curtis}(1938)}]{cole1938}%
  \BibitemOpen
  \bibfield  {author} {\bibinfo {author} {\bibfnamefont {K.~S.}\ \bibnamefont {Cole}}\ and\ \bibinfo {author} {\bibfnamefont {H.~J.}\ \bibnamefont {Curtis}},\ }\bibfield  {title} {\bibinfo {title} {Electric impedance of nitella during activity},\ }\href@noop {} {\bibfield  {journal} {\bibinfo  {journal} {Journal of General Physiology}\ }\textbf {\bibinfo {volume} {22}},\ \bibinfo {pages} {37} (\bibinfo {year} {1938})}\BibitemShut {NoStop}%
\bibitem [{\citenamefont {Hodgkin}\ and\ \citenamefont {Rushton}(1946)}]{hodgkin1946}%
  \BibitemOpen
  \bibfield  {author} {\bibinfo {author} {\bibfnamefont {A.~L.}\ \bibnamefont {Hodgkin}}\ and\ \bibinfo {author} {\bibfnamefont {W.~A.~H.}\ \bibnamefont {Rushton}},\ }\href@noop {} {\bibfield  {journal} {\bibinfo  {journal} {Proceedings of the Royal Society of London. Series B - Biological Sciences}\ }\textbf {\bibinfo {volume} {133}},\ \bibinfo {pages} {444–479} (\bibinfo {year} {1946})}\BibitemShut {NoStop}%
\bibitem [{\citenamefont {Rall}(1962)}]{rall1962}%
  \BibitemOpen
  \bibfield  {author} {\bibinfo {author} {\bibfnamefont {W.}~\bibnamefont {Rall}},\ }\bibfield  {title} {\bibinfo {title} {Theory of physiological properties of dendrites},\ }\href@noop {} {\bibfield  {journal} {\bibinfo  {journal} {Annals of the New York Academy of Sciences}\ }\textbf {\bibinfo {volume} {96}},\ \bibinfo {pages} {1071} (\bibinfo {year} {1962})}\BibitemShut {NoStop}%
\bibitem [{\citenamefont {Dayan}\ and\ \citenamefont {Abbott}(2001)}]{dayan2001}%
  \BibitemOpen
  \bibfield  {author} {\bibinfo {author} {\bibfnamefont {P.}~\bibnamefont {Dayan}}\ and\ \bibinfo {author} {\bibfnamefont {L.~F.}\ \bibnamefont {Abbott}},\ }\href@noop {} {\emph {\bibinfo {title} {Theoretical Neuroscience}}},\ Computational Neuroscience\ (\bibinfo  {publisher} {MIT Press},\ \bibinfo {address} {London, England},\ \bibinfo {year} {2001})\BibitemShut {NoStop}%
\bibitem [{\citenamefont {Savtchenko}\ \emph {et~al.}(2017)\citenamefont {Savtchenko}, \citenamefont {Poo},\ and\ \citenamefont {Rusakov}}]{savtchenko2017electrodiffusion}%
  \BibitemOpen
  \bibfield  {author} {\bibinfo {author} {\bibfnamefont {L.~P.}\ \bibnamefont {Savtchenko}}, \bibinfo {author} {\bibfnamefont {M.~M.}\ \bibnamefont {Poo}},\ and\ \bibinfo {author} {\bibfnamefont {D.~A.}\ \bibnamefont {Rusakov}},\ }\bibfield  {title} {\bibinfo {title} {Electrodiffusion phenomena in neuroscience: a neglected companion},\ }\href@noop {} {\bibfield  {journal} {\bibinfo  {journal} {Nature reviews Neuroscience}\ }\textbf {\bibinfo {volume} {18}},\ \bibinfo {pages} {598} (\bibinfo {year} {2017})}\BibitemShut {NoStop}%
\bibitem [{\citenamefont {Sylantyev}\ \emph {et~al.}(2008)\citenamefont {Sylantyev}, \citenamefont {Savtchenko}, \citenamefont {Niu}, \citenamefont {Ivanov}, \citenamefont {Jensen}, \citenamefont {Kullmann}, \citenamefont {Xiao},\ and\ \citenamefont {Rusakov}}]{sylantyev2008}%
  \BibitemOpen
  \bibfield  {author} {\bibinfo {author} {\bibfnamefont {S.}~\bibnamefont {Sylantyev}}, \bibinfo {author} {\bibfnamefont {L.~P.}\ \bibnamefont {Savtchenko}}, \bibinfo {author} {\bibfnamefont {Y.-P.}\ \bibnamefont {Niu}}, \bibinfo {author} {\bibfnamefont {A.~I.}\ \bibnamefont {Ivanov}}, \bibinfo {author} {\bibfnamefont {T.~P.}\ \bibnamefont {Jensen}}, \bibinfo {author} {\bibfnamefont {D.~M.}\ \bibnamefont {Kullmann}}, \bibinfo {author} {\bibfnamefont {M.-Y.}\ \bibnamefont {Xiao}},\ and\ \bibinfo {author} {\bibfnamefont {D.~A.}\ \bibnamefont {Rusakov}},\ }\bibfield  {title} {\bibinfo {title} {Electric fields due to synaptic currents sharpen excitatory transmission},\ }\href@noop {} {\bibfield  {journal} {\bibinfo  {journal} {Science}\ }\textbf {\bibinfo {volume} {319}},\ \bibinfo {pages} {1845} (\bibinfo {year} {2008})}\BibitemShut {NoStop}%
\bibitem [{\citenamefont {Cartailler}\ \emph {et~al.}(2018)\citenamefont {Cartailler}, \citenamefont {Kwon}, \citenamefont {Yuste},\ and\ \citenamefont {Holcman}}]{cartailler2018deconvolution}%
  \BibitemOpen
  \bibfield  {author} {\bibinfo {author} {\bibfnamefont {J.}~\bibnamefont {Cartailler}}, \bibinfo {author} {\bibfnamefont {T.}~\bibnamefont {Kwon}}, \bibinfo {author} {\bibfnamefont {R.}~\bibnamefont {Yuste}},\ and\ \bibinfo {author} {\bibfnamefont {D.}~\bibnamefont {Holcman}},\ }\bibfield  {title} {\bibinfo {title} {Deconvolution of voltage sensor time series and electro-diffusion modeling reveal the role of spine geometry in controlling synaptic strength},\ }\href@noop {} {\bibfield  {journal} {\bibinfo  {journal} {Neuron}\ }\textbf {\bibinfo {volume} {97}},\ \bibinfo {pages} {1126} (\bibinfo {year} {2018})}\BibitemShut {NoStop}%
\bibitem [{\citenamefont {Huang}\ and\ \citenamefont {Levitt}(1977)}]{huang1977theoretical}%
  \BibitemOpen
  \bibfield  {author} {\bibinfo {author} {\bibfnamefont {W.-T.}\ \bibnamefont {Huang}}\ and\ \bibinfo {author} {\bibfnamefont {D.}~\bibnamefont {Levitt}},\ }\bibfield  {title} {\bibinfo {title} {Theoretical calculation of the dielectric constant of a bilayer membrane},\ }\href@noop {} {\bibfield  {journal} {\bibinfo  {journal} {Biophysical Journal}\ }\textbf {\bibinfo {volume} {17}},\ \bibinfo {pages} {111} (\bibinfo {year} {1977})}\BibitemShut {NoStop}%
\bibitem [{\citenamefont {Nymeyer}\ and\ \citenamefont {Zhou}(2008)}]{nymeyer2008method}%
  \BibitemOpen
  \bibfield  {author} {\bibinfo {author} {\bibfnamefont {H.}~\bibnamefont {Nymeyer}}\ and\ \bibinfo {author} {\bibfnamefont {H.-X.}\ \bibnamefont {Zhou}},\ }\bibfield  {title} {\bibinfo {title} {A method to determine dielectric constants in nonhomogeneous systems: application to biological membranes},\ }\href@noop {} {\bibfield  {journal} {\bibinfo  {journal} {Biophysical Journal}\ }\textbf {\bibinfo {volume} {94}},\ \bibinfo {pages} {1185} (\bibinfo {year} {2008})}\BibitemShut {NoStop}%
\bibitem [{\citenamefont {Nernst}(1888)}]{Nernst1888}%
  \BibitemOpen
  \bibfield  {author} {\bibinfo {author} {\bibfnamefont {W.}~\bibnamefont {Nernst}},\ }\bibfield  {title} {\bibinfo {title} {Zur kinetik der in lösung befindlichen körper},\ }\href@noop {} {\bibfield  {journal} {\bibinfo  {journal} {Zeitschrift für Physikalische Chemie}\ }\textbf {\bibinfo {volume} {2U}},\ \bibinfo {pages} {613} (\bibinfo {year} {1888})}\BibitemShut {NoStop}%
\bibitem [{\citenamefont {Nernst}(1889)}]{Nernst1889}%
  \BibitemOpen
  \bibfield  {author} {\bibinfo {author} {\bibfnamefont {W.}~\bibnamefont {Nernst}},\ }\bibfield  {title} {\bibinfo {title} {Die elektromotorische wirksamkeit der jonen},\ }\href@noop {} {\bibfield  {journal} {\bibinfo  {journal} {Zeitschrift für Physikalische Chemie}\ }\textbf {\bibinfo {volume} {4U}},\ \bibinfo {pages} {129} (\bibinfo {year} {1889})}\BibitemShut {NoStop}%
\bibitem [{\citenamefont {Planck}(1890)}]{Planck1890}%
  \BibitemOpen
  \bibfield  {author} {\bibinfo {author} {\bibfnamefont {M.}~\bibnamefont {Planck}},\ }\bibfield  {title} {\bibinfo {title} {Ueber die potentialdifferenz zwischen zwei verdünnten lösungen binärer electrolyte},\ }\href@noop {} {\bibfield  {journal} {\bibinfo  {journal} {Annalen der Physik}\ }\textbf {\bibinfo {volume} {276}},\ \bibinfo {pages} {561} (\bibinfo {year} {1890})}\BibitemShut {NoStop}%
\bibitem [{\citenamefont {Bazant}\ \emph {et~al.}(2004)\citenamefont {Bazant}, \citenamefont {Thornton},\ and\ \citenamefont {Ajdari}}]{Bazant04}%
  \BibitemOpen
  \bibfield  {author} {\bibinfo {author} {\bibfnamefont {M.~Z.}\ \bibnamefont {Bazant}}, \bibinfo {author} {\bibfnamefont {K.}~\bibnamefont {Thornton}},\ and\ \bibinfo {author} {\bibfnamefont {A.}~\bibnamefont {Ajdari}},\ }\bibfield  {title} {\bibinfo {title} {Diffuse-charge dynamics in electrochemical systems},\ }\href@noop {} {\bibfield  {journal} {\bibinfo  {journal} {Physical Review E}\ }\textbf {\bibinfo {volume} {70}},\ \bibinfo {pages} {021506} (\bibinfo {year} {2004})}\BibitemShut {NoStop}%
\bibitem [{\citenamefont {Fong}\ \emph {et~al.}(2020)\citenamefont {Fong}, \citenamefont {Bergstrom}, \citenamefont {McCloskey},\ and\ \citenamefont {Mandadapu}}]{fong2020transport}%
  \BibitemOpen
  \bibfield  {author} {\bibinfo {author} {\bibfnamefont {K.~D.}\ \bibnamefont {Fong}}, \bibinfo {author} {\bibfnamefont {H.~K.}\ \bibnamefont {Bergstrom}}, \bibinfo {author} {\bibfnamefont {B.~D.}\ \bibnamefont {McCloskey}},\ and\ \bibinfo {author} {\bibfnamefont {K.~K.}\ \bibnamefont {Mandadapu}},\ }\bibfield  {title} {\bibinfo {title} {Transport phenomena in electrolyte solutions: Nonequilibrium thermodynamics and statistical mechanics},\ }\href@noop {} {\bibfield  {journal} {\bibinfo  {journal} {AIChE Journal}\ }\textbf {\bibinfo {volume} {66}},\ \bibinfo {pages} {e17091} (\bibinfo {year} {2020})}\BibitemShut {NoStop}%
\bibitem [{\citenamefont {Tedesco}\ \emph {et~al.}(2016)\citenamefont {Tedesco}, \citenamefont {Hamelers},\ and\ \citenamefont {Biesheuvel}}]{tedesco2016nernst}%
  \BibitemOpen
  \bibfield  {author} {\bibinfo {author} {\bibfnamefont {M.}~\bibnamefont {Tedesco}}, \bibinfo {author} {\bibfnamefont {H.}~\bibnamefont {Hamelers}},\ and\ \bibinfo {author} {\bibfnamefont {P.}~\bibnamefont {Biesheuvel}},\ }\bibfield  {title} {\bibinfo {title} {Nernst-planck transport theory for (reverse) electrodialysis: I. effect of co-ion transport through the membranes},\ }\href@noop {} {\bibfield  {journal} {\bibinfo  {journal} {Journal of Membrane Science}\ }\textbf {\bibinfo {volume} {510}},\ \bibinfo {pages} {370} (\bibinfo {year} {2016})}\BibitemShut {NoStop}%
\bibitem [{\citenamefont {Mei}\ \emph {et~al.}(2018)\citenamefont {Mei}, \citenamefont {Munteshari}, \citenamefont {Lau}, \citenamefont {Dunn},\ and\ \citenamefont {Pilon}}]{mei2018physical}%
  \BibitemOpen
  \bibfield  {author} {\bibinfo {author} {\bibfnamefont {B.-A.}\ \bibnamefont {Mei}}, \bibinfo {author} {\bibfnamefont {O.}~\bibnamefont {Munteshari}}, \bibinfo {author} {\bibfnamefont {J.}~\bibnamefont {Lau}}, \bibinfo {author} {\bibfnamefont {B.}~\bibnamefont {Dunn}},\ and\ \bibinfo {author} {\bibfnamefont {L.}~\bibnamefont {Pilon}},\ }\bibfield  {title} {\bibinfo {title} {Physical interpretations of nyquist plots for edlc electrodes and devices},\ }\href@noop {} {\bibfield  {journal} {\bibinfo  {journal} {The Journal of Physical Chemistry C}\ }\textbf {\bibinfo {volume} {122}},\ \bibinfo {pages} {194} (\bibinfo {year} {2018})}\BibitemShut {NoStop}%
\bibitem [{\citenamefont {Wu}(2022)}]{wu2022understanding}%
  \BibitemOpen
  \bibfield  {author} {\bibinfo {author} {\bibfnamefont {J.}~\bibnamefont {Wu}},\ }\bibfield  {title} {\bibinfo {title} {Understanding the electric double-layer structure, capacitance, and charging dynamics},\ }\href@noop {} {\bibfield  {journal} {\bibinfo  {journal} {Chemical Reviews}\ }\textbf {\bibinfo {volume} {122}},\ \bibinfo {pages} {10821} (\bibinfo {year} {2022})}\BibitemShut {NoStop}%
\bibitem [{\citenamefont {Chen}\ \emph {et~al.}(1997)\citenamefont {Chen}, \citenamefont {Lear},\ and\ \citenamefont {Eisenberg}}]{chen1997}%
  \BibitemOpen
  \bibfield  {author} {\bibinfo {author} {\bibfnamefont {D.}~\bibnamefont {Chen}}, \bibinfo {author} {\bibfnamefont {J.}~\bibnamefont {Lear}},\ and\ \bibinfo {author} {\bibfnamefont {B.}~\bibnamefont {Eisenberg}},\ }\bibfield  {title} {\bibinfo {title} {Permeation through an open channel: Poisson-nernst-planck theory of a synthetic ionic channel},\ }\href@noop {} {\bibfield  {journal} {\bibinfo  {journal} {Biophysical Journal}\ }\textbf {\bibinfo {volume} {72}},\ \bibinfo {pages} {97} (\bibinfo {year} {1997})}\BibitemShut {NoStop}%
\bibitem [{\citenamefont {Liu}\ and\ \citenamefont {Eisenberg}(2014)}]{liu2014}%
  \BibitemOpen
  \bibfield  {author} {\bibinfo {author} {\bibfnamefont {J.-L.}\ \bibnamefont {Liu}}\ and\ \bibinfo {author} {\bibfnamefont {B.}~\bibnamefont {Eisenberg}},\ }\bibfield  {title} {\bibinfo {title} {{Poisson-Nernst-Planck-Fermi theory for modeling biological ion channels}},\ }\href@noop {} {\bibfield  {journal} {\bibinfo  {journal} {The Journal of Chemical Physics}\ }\textbf {\bibinfo {volume} {141}},\ \bibinfo {pages} {22D532} (\bibinfo {year} {2014})}\BibitemShut {NoStop}%
\bibitem [{\citenamefont {Singer}\ \emph {et~al.}(2008)\citenamefont {Singer}, \citenamefont {Gillespie}, \citenamefont {Norbury},\ and\ \citenamefont {Eisenberg}}]{singer2008}%
  \BibitemOpen
  \bibfield  {author} {\bibinfo {author} {\bibfnamefont {A.}~\bibnamefont {Singer}}, \bibinfo {author} {\bibfnamefont {D.}~\bibnamefont {Gillespie}}, \bibinfo {author} {\bibfnamefont {J.}~\bibnamefont {Norbury}},\ and\ \bibinfo {author} {\bibfnamefont {R.~S.}\ \bibnamefont {Eisenberg}},\ }\bibfield  {title} {\bibinfo {title} {Singular perturbation analysis of the steady-state poisson–nernst–planck system: Applications to ion channels},\ }\href@noop {} {\bibfield  {journal} {\bibinfo  {journal} {European Journal of Applied Mathematics}\ }\textbf {\bibinfo {volume} {19}},\ \bibinfo {pages} {541–560} (\bibinfo {year} {2008})}\BibitemShut {NoStop}%
\bibitem [{\citenamefont {Savtchenko}\ \emph {et~al.}(2004)\citenamefont {Savtchenko}, \citenamefont {Kulahin}, \citenamefont {Korogod},\ and\ \citenamefont {Rusakov}}]{savtchenko2004electric}%
  \BibitemOpen
  \bibfield  {author} {\bibinfo {author} {\bibfnamefont {L.~P.}\ \bibnamefont {Savtchenko}}, \bibinfo {author} {\bibfnamefont {N.}~\bibnamefont {Kulahin}}, \bibinfo {author} {\bibfnamefont {S.~M.}\ \bibnamefont {Korogod}},\ and\ \bibinfo {author} {\bibfnamefont {D.~A.}\ \bibnamefont {Rusakov}},\ }\bibfield  {title} {\bibinfo {title} {Electric fields of synaptic currents could influence diffusion of charged neurotransmitter molecules},\ }\href@noop {} {\bibfield  {journal} {\bibinfo  {journal} {Synapse}\ }\textbf {\bibinfo {volume} {51}},\ \bibinfo {pages} {270} (\bibinfo {year} {2004})}\BibitemShut {NoStop}%
\bibitem [{\citenamefont {Cartailler}\ \emph {et~al.}(2017)\citenamefont {Cartailler}, \citenamefont {Schuss},\ and\ \citenamefont {Holcman}}]{cartailler2017}%
  \BibitemOpen
  \bibfield  {author} {\bibinfo {author} {\bibfnamefont {J.}~\bibnamefont {Cartailler}}, \bibinfo {author} {\bibfnamefont {Z.}~\bibnamefont {Schuss}},\ and\ \bibinfo {author} {\bibfnamefont {D.}~\bibnamefont {Holcman}},\ }\bibfield  {title} {\bibinfo {title} {Analysis of the poisson–nernst–planck equation in a ball for modeling the voltage–current relation in neurobiological microdomains},\ }\href@noop {} {\bibfield  {journal} {\bibinfo  {journal} {Physica D: Nonlinear Phenomena}\ }\textbf {\bibinfo {volume} {339}},\ \bibinfo {pages} {39} (\bibinfo {year} {2017})}\BibitemShut {NoStop}%
\bibitem [{\citenamefont {Lagache}\ \emph {et~al.}(2019)\citenamefont {Lagache}, \citenamefont {Jayant},\ and\ \citenamefont {Yuste}}]{lagache2019}%
  \BibitemOpen
  \bibfield  {author} {\bibinfo {author} {\bibfnamefont {T.}~\bibnamefont {Lagache}}, \bibinfo {author} {\bibfnamefont {K.}~\bibnamefont {Jayant}},\ and\ \bibinfo {author} {\bibfnamefont {R.}~\bibnamefont {Yuste}},\ }\bibfield  {title} {\bibinfo {title} {Electrodiffusion models of synaptic potentials in dendritic spines},\ }\href@noop {} {\bibfield  {journal} {\bibinfo  {journal} {Journal of Computational Neuroscience}\ }\textbf {\bibinfo {volume} {47}},\ \bibinfo {pages} {77–89} (\bibinfo {year} {2019})}\BibitemShut {NoStop}%
\bibitem [{\citenamefont {Pods}\ \emph {et~al.}(2013)\citenamefont {Pods}, \citenamefont {Sch{\"o}nke},\ and\ \citenamefont {Bastian}}]{pods2013electrodiffusion}%
  \BibitemOpen
  \bibfield  {author} {\bibinfo {author} {\bibfnamefont {J.}~\bibnamefont {Pods}}, \bibinfo {author} {\bibfnamefont {J.}~\bibnamefont {Sch{\"o}nke}},\ and\ \bibinfo {author} {\bibfnamefont {P.}~\bibnamefont {Bastian}},\ }\bibfield  {title} {\bibinfo {title} {Electrodiffusion models of neurons and extracellular space using the poisson-nernst-planck equations—numerical simulation of the intra-and extracellular potential for an axon model},\ }\href@noop {} {\bibfield  {journal} {\bibinfo  {journal} {Biophysical Journal}\ }\textbf {\bibinfo {volume} {105}},\ \bibinfo {pages} {242} (\bibinfo {year} {2013})}\BibitemShut {NoStop}%
\bibitem [{\citenamefont {Gulati}\ and\ \citenamefont {Rudraraju}(2023)}]{gulati2023}%
  \BibitemOpen
  \bibfield  {author} {\bibinfo {author} {\bibfnamefont {R.}~\bibnamefont {Gulati}}\ and\ \bibinfo {author} {\bibfnamefont {S.}~\bibnamefont {Rudraraju}},\ }\bibfield  {title} {\bibinfo {title} {Spatio-temporal modeling of saltatory conduction in neurons using poisson–nernst–planck treatment and estimation of conduction velocity},\ }\href@noop {} {\bibfield  {journal} {\bibinfo  {journal} {Brain Multiphysics}\ }\textbf {\bibinfo {volume} {4}},\ \bibinfo {pages} {100061} (\bibinfo {year} {2023})}\BibitemShut {NoStop}%
\bibitem [{\citenamefont {Jackson}(1999)}]{jackson1999classical}%
  \BibitemOpen
  \bibfield  {author} {\bibinfo {author} {\bibfnamefont {J.~D.}\ \bibnamefont {Jackson}},\ }\href@noop {} {\emph {\bibinfo {title} {Classical Electrodynamics}}}\ (\bibinfo  {publisher} {John Wiley \& Sons},\ \bibinfo {year} {1999})\BibitemShut {NoStop}%
\bibitem [{\citenamefont {Kovetz}(200)}]{Kovetz2000}%
  \BibitemOpen
  \bibfield  {author} {\bibinfo {author} {\bibfnamefont {A.}~\bibnamefont {Kovetz}},\ }\href@noop {} {\emph {\bibinfo {title} {Electromagnetic Theory}}}\ (\bibinfo  {publisher} {Clarendon Press},\ \bibinfo {address} {Oxford, England},\ \bibinfo {year} {200})\BibitemShut {NoStop}%
\bibitem [{\citenamefont {Volkov}\ and\ \citenamefont {Hampton}(2008)}]{volkov2008energetics}%
  \BibitemOpen
  \bibfield  {author} {\bibinfo {author} {\bibfnamefont {A.~G.}\ \bibnamefont {Volkov}}\ and\ \bibinfo {author} {\bibfnamefont {T.}~\bibnamefont {Hampton}},\ }\bibfield  {title} {\bibinfo {title} {Energetics of membrane permeability},\ }\href@noop {} {\bibfield  {journal} {\bibinfo  {journal} {Advances in Planar Lipid Bilayers and Liposomes}\ }\textbf {\bibinfo {volume} {8}},\ \bibinfo {pages} {155} (\bibinfo {year} {2008})}\BibitemShut {NoStop}%
\bibitem [{\citenamefont {Skou}(1957)}]{SKOU1957394}%
  \BibitemOpen
  \bibfield  {author} {\bibinfo {author} {\bibfnamefont {J.~C.}\ \bibnamefont {Skou}},\ }\bibfield  {title} {\bibinfo {title} {The influence of some cations on an adenosine triphosphatase from peripheral nerves},\ }\href@noop {} {\bibfield  {journal} {\bibinfo  {journal} {Biochimica et Biophysica Acta}\ }\textbf {\bibinfo {volume} {23}},\ \bibinfo {pages} {394} (\bibinfo {year} {1957})}\BibitemShut {NoStop}%
\bibitem [{\citenamefont {Debye}\ and\ \citenamefont {H{\"u}ckel}(1923)}]{debye1923}%
  \BibitemOpen
  \bibfield  {author} {\bibinfo {author} {\bibfnamefont {P.}~\bibnamefont {Debye}}\ and\ \bibinfo {author} {\bibfnamefont {E.}~\bibnamefont {H{\"u}ckel}},\ }\bibfield  {title} {\bibinfo {title} {Zur theorie der elektrolyte i},\ }\href@noop {} {\bibfield  {journal} {\bibinfo  {journal} {Physikalische Zeitschrift}\ }\textbf {\bibinfo {volume} {24}},\ \bibinfo {pages} {185} (\bibinfo {year} {1923})}\BibitemShut {NoStop}%
\bibitem [{\citenamefont {Gouy}(1910)}]{Gouy1910}%
  \BibitemOpen
  \bibfield  {author} {\bibinfo {author} {\bibfnamefont {M.}~\bibnamefont {Gouy}},\ }\bibfield  {title} {\bibinfo {title} {Sur la constitution de la charge électrique à la surface d’un électrolyte},\ }\href@noop {} {\bibfield  {journal} {\bibinfo  {journal} {Journal de Physique Théorique et Appliquée}\ }\textbf {\bibinfo {volume} {9}},\ \bibinfo {pages} {457–468} (\bibinfo {year} {1910})}\BibitemShut {NoStop}%
\bibitem [{\citenamefont {Van{\'{y}}sek}(2015)}]{vanysek2015}%
  \BibitemOpen
  \bibfield  {author} {\bibinfo {author} {\bibfnamefont {P.}~\bibnamefont {Van{\'{y}}sek}},\ }\bibinfo {title} {Ionic conductivity and diffusion at infinite dilution}\ (\bibinfo  {publisher} {CRC Handbook of Chemistry and Physics, 95th Edition, CRC Press, Taylor \& Francis Group, Boca Raton, FL},\ \bibinfo {year} {2015})\BibitemShut {NoStop}%
\bibitem [{\citenamefont {Hughes}(2012)}]{hughes2012finite}%
  \BibitemOpen
  \bibfield  {author} {\bibinfo {author} {\bibfnamefont {T.~J.}\ \bibnamefont {Hughes}},\ }\href@noop {} {\emph {\bibinfo {title} {The Finite Element Method: Linear Static and Dynamic Finite Element Analysis}}}\ (\bibinfo  {publisher} {Courier Corporation},\ \bibinfo {year} {2012})\BibitemShut {NoStop}%
\bibitem [{\citenamefont {Reddy}(2015)}]{reddy2015introduction}%
  \BibitemOpen
  \bibfield  {author} {\bibinfo {author} {\bibfnamefont {J.~N.}\ \bibnamefont {Reddy}},\ }\href@noop {} {\emph {\bibinfo {title} {An Introduction to Nonlinear Finite Element Analysis: with Applications to Heat Transfer, Fluid Mechanics, and Solid Mechanics}}}\ (\bibinfo  {publisher} {Oxford University Press},\ \bibinfo {year} {2015})\BibitemShut {NoStop}%
\bibitem [{\citenamefont {Papadopoulos}(2015)}]{papadopoulos2015fem}%
  \BibitemOpen
  \bibfield  {author} {\bibinfo {author} {\bibfnamefont {P.}~\bibnamefont {Papadopoulos}},\ }\href@noop {} {\emph {\bibinfo {title} {Introduction to the Finite Element Method}}}\ (\bibinfo  {publisher} {University of California, Berkeley},\ \bibinfo {year} {2015})\BibitemShut {NoStop}%
\bibitem [{\citenamefont {Scroggs}\ \emph {et~al.}(2022{\natexlab{a}})\citenamefont {Scroggs}, \citenamefont {Dokken}, \citenamefont {Richardson},\ and\ \citenamefont {Wells}}]{Scroggs22a}%
  \BibitemOpen
  \bibfield  {author} {\bibinfo {author} {\bibfnamefont {M.~W.}\ \bibnamefont {Scroggs}}, \bibinfo {author} {\bibfnamefont {J.~S.}\ \bibnamefont {Dokken}}, \bibinfo {author} {\bibfnamefont {C.~N.}\ \bibnamefont {Richardson}},\ and\ \bibinfo {author} {\bibfnamefont {G.~N.}\ \bibnamefont {Wells}},\ }\bibfield  {title} {\bibinfo {title} {Construction of arbitrary order finite element degree-of-freedom maps on polygonal and polyhedral cell meshes},\ }\href@noop {} {\bibfield  {journal} {\bibinfo  {journal} {ACM Transactions on Mathematical Software}\ }\textbf {\bibinfo {volume} {48}},\ \bibinfo {pages} {1} (\bibinfo {year} {2022}{\natexlab{a}})}\BibitemShut {NoStop}%
\bibitem [{\citenamefont {Scroggs}\ \emph {et~al.}(2022{\natexlab{b}})\citenamefont {Scroggs}, \citenamefont {Baratta}, \citenamefont {Richardson},\ and\ \citenamefont {Wells}}]{Scroggs22b}%
  \BibitemOpen
  \bibfield  {author} {\bibinfo {author} {\bibfnamefont {M.~W.}\ \bibnamefont {Scroggs}}, \bibinfo {author} {\bibfnamefont {I.~A.}\ \bibnamefont {Baratta}}, \bibinfo {author} {\bibfnamefont {C.~N.}\ \bibnamefont {Richardson}},\ and\ \bibinfo {author} {\bibfnamefont {G.~N.}\ \bibnamefont {Wells}},\ }\bibfield  {title} {\bibinfo {title} {Basix: a runtime finite element basis evaluation library},\ }\href@noop {} {\bibfield  {journal} {\bibinfo  {journal} {Journal of Open Source Software}\ }\textbf {\bibinfo {volume} {7}},\ \bibinfo {pages} {3982} (\bibinfo {year} {2022}{\natexlab{b}})}\BibitemShut {NoStop}%
\bibitem [{\citenamefont {Aln\ae{}s}\ \emph {et~al.}(2014)\citenamefont {Aln\ae{}s}, \citenamefont {Logg}, \citenamefont {\O{}lgaard}, \citenamefont {Rognes},\ and\ \citenamefont {Wells}}]{Alnaes14}%
  \BibitemOpen
  \bibfield  {author} {\bibinfo {author} {\bibfnamefont {M.~S.}\ \bibnamefont {Aln\ae{}s}}, \bibinfo {author} {\bibfnamefont {A.}~\bibnamefont {Logg}}, \bibinfo {author} {\bibfnamefont {K.~B.}\ \bibnamefont {\O{}lgaard}}, \bibinfo {author} {\bibfnamefont {M.~E.}\ \bibnamefont {Rognes}},\ and\ \bibinfo {author} {\bibfnamefont {G.~N.}\ \bibnamefont {Wells}},\ }\bibfield  {title} {\bibinfo {title} {Unified form language: A domain-specific language for weak formulations of partial differential equations},\ }\href@noop {} {\bibfield  {journal} {\bibinfo  {journal} {ACM Transactions on Mathematical Software}\ }\textbf {\bibinfo {volume} {40}},\ \bibinfo {pages} {1} (\bibinfo {year} {2014})}\BibitemShut {NoStop}%
\bibitem [{\citenamefont {Geuzaine}\ and\ \citenamefont {Remacle}(2009)}]{Geuzaine09}%
  \BibitemOpen
  \bibfield  {author} {\bibinfo {author} {\bibfnamefont {C.}~\bibnamefont {Geuzaine}}\ and\ \bibinfo {author} {\bibfnamefont {J.-F.}\ \bibnamefont {Remacle}},\ }\bibfield  {title} {\bibinfo {title} {Gmsh: A 3-d finite element mesh generator with built-in pre- and post-processing facilities},\ }\href@noop {} {\bibfield  {journal} {\bibinfo  {journal} {International Journal for Numerical Methods in Engineering}\ }\textbf {\bibinfo {volume} {79}},\ \bibinfo {pages} {1309} (\bibinfo {year} {2009})}\BibitemShut {NoStop}%
\bibitem [{\citenamefont {Hodgkin}\ and\ \citenamefont {Huxley}(1939)}]{hodgkin1939action}%
  \BibitemOpen
  \bibfield  {author} {\bibinfo {author} {\bibfnamefont {A.~L.}\ \bibnamefont {Hodgkin}}\ and\ \bibinfo {author} {\bibfnamefont {A.~F.}\ \bibnamefont {Huxley}},\ }\bibfield  {title} {\bibinfo {title} {Action potentials recorded from inside a nerve fibre},\ }\href@noop {} {\bibfield  {journal} {\bibinfo  {journal} {Nature}\ }\textbf {\bibinfo {volume} {144}},\ \bibinfo {pages} {710} (\bibinfo {year} {1939})}\BibitemShut {NoStop}%
\bibitem [{\citenamefont {Chapman}(1913)}]{chapman1913li}%
  \BibitemOpen
  \bibfield  {author} {\bibinfo {author} {\bibfnamefont {D.~L.}\ \bibnamefont {Chapman}},\ }\bibfield  {title} {\bibinfo {title} {L{I}. {A} contribution to the theory of electrocapillarity},\ }\href@noop {} {\bibfield  {journal} {\bibinfo  {journal} {The London, Edinburgh, and Dublin Philosophical Magazine and Journal of Science}\ }\textbf {\bibinfo {volume} {25}},\ \bibinfo {pages} {475} (\bibinfo {year} {1913})}\BibitemShut {NoStop}%
\bibitem [{\citenamefont {Moldenhauer}\ \emph {et~al.}(2016)\citenamefont {Moldenhauer}, \citenamefont {D{\'\i}az-Franulic}, \citenamefont {Gonz{\'a}lez-Nilo},\ and\ \citenamefont {Naranjo}}]{moldenhauer2016effective}%
  \BibitemOpen
  \bibfield  {author} {\bibinfo {author} {\bibfnamefont {H.}~\bibnamefont {Moldenhauer}}, \bibinfo {author} {\bibfnamefont {I.}~\bibnamefont {D{\'\i}az-Franulic}}, \bibinfo {author} {\bibfnamefont {F.}~\bibnamefont {Gonz{\'a}lez-Nilo}},\ and\ \bibinfo {author} {\bibfnamefont {D.}~\bibnamefont {Naranjo}},\ }\bibfield  {title} {\bibinfo {title} {Effective pore size and radius of capture for k+ ions in k-channels},\ }\href@noop {} {\bibfield  {journal} {\bibinfo  {journal} {Scientific Reports}\ }\textbf {\bibinfo {volume} {6}},\ \bibinfo {pages} {19893} (\bibinfo {year} {2016})}\BibitemShut {NoStop}%
\bibitem [{\citenamefont {Rosenbluth}(1976)}]{rosenbluth1976intramembranous}%
  \BibitemOpen
  \bibfield  {author} {\bibinfo {author} {\bibfnamefont {J.}~\bibnamefont {Rosenbluth}},\ }\bibfield  {title} {\bibinfo {title} {Intramembranous particle distribution at the node of ranvier and adjacent axolemma in myelinated axons of the frog brain},\ }\href@noop {} {\bibfield  {journal} {\bibinfo  {journal} {Journal of Neurocytology}\ }\textbf {\bibinfo {volume} {5}},\ \bibinfo {pages} {731} (\bibinfo {year} {1976})}\BibitemShut {NoStop}%
\bibitem [{\citenamefont {Neher}\ and\ \citenamefont {Sakmann}(1976)}]{neher1976single}%
  \BibitemOpen
  \bibfield  {author} {\bibinfo {author} {\bibfnamefont {E.}~\bibnamefont {Neher}}\ and\ \bibinfo {author} {\bibfnamefont {B.}~\bibnamefont {Sakmann}},\ }\bibfield  {title} {\bibinfo {title} {Single-channel currents recorded from membrane of denervated frog muscle fibres},\ }\href@noop {} {\bibfield  {journal} {\bibinfo  {journal} {Nature}\ }\textbf {\bibinfo {volume} {260}},\ \bibinfo {pages} {799} (\bibinfo {year} {1976})}\BibitemShut {NoStop}%
\bibitem [{\citenamefont {Neher}(1992)}]{neher1992ion}%
  \BibitemOpen
  \bibfield  {author} {\bibinfo {author} {\bibfnamefont {E.}~\bibnamefont {Neher}},\ }\bibfield  {title} {\bibinfo {title} {Ion channels for communication between and within cells},\ }\href@noop {} {\bibfield  {journal} {\bibinfo  {journal} {Neuron}\ }\textbf {\bibinfo {volume} {8}},\ \bibinfo {pages} {605} (\bibinfo {year} {1992})}\BibitemShut {NoStop}%
\bibitem [{\citenamefont {Hodgkin}\ \emph {et~al.}(1952)\citenamefont {Hodgkin}, \citenamefont {Huxley},\ and\ \citenamefont {Katz}}]{hodgkin1952measurement}%
  \BibitemOpen
  \bibfield  {author} {\bibinfo {author} {\bibfnamefont {A.~L.}\ \bibnamefont {Hodgkin}}, \bibinfo {author} {\bibfnamefont {A.~F.}\ \bibnamefont {Huxley}},\ and\ \bibinfo {author} {\bibfnamefont {B.}~\bibnamefont {Katz}},\ }\bibfield  {title} {\bibinfo {title} {Measurement of current-voltage relations in the membrane of the giant axon of loligo},\ }\href@noop {} {\bibfield  {journal} {\bibinfo  {journal} {The Journal of Physiology}\ }\textbf {\bibinfo {volume} {116}},\ \bibinfo {pages} {424} (\bibinfo {year} {1952})}\BibitemShut {NoStop}%
\bibitem [{Note1()}]{Note1}%
  \BibitemOpen
  \bibinfo {note} {Although reminiscent of self-similarity, standard methods of identifying similarity solutions fail here as the near-field portion is not self-similar \textcolor {magenta}{(Fig.~S6)}.}\BibitemShut {Stop}%
\bibitem [{\citenamefont {Sahu}\ \emph {et~al.}(2020)\citenamefont {Sahu}, \citenamefont {Glisman}, \citenamefont {Tchoufag},\ and\ \citenamefont {Mandadapu}}]{sahu2020geometry}%
  \BibitemOpen
  \bibfield  {author} {\bibinfo {author} {\bibfnamefont {A.}~\bibnamefont {Sahu}}, \bibinfo {author} {\bibfnamefont {A.}~\bibnamefont {Glisman}}, \bibinfo {author} {\bibfnamefont {J.}~\bibnamefont {Tchoufag}},\ and\ \bibinfo {author} {\bibfnamefont {K.~K.}\ \bibnamefont {Mandadapu}},\ }\bibfield  {title} {\bibinfo {title} {Geometry and dynamics of lipid membranes: the scriven-love number},\ }\href@noop {} {\bibfield  {journal} {\bibinfo  {journal} {Physical Review E}\ }\textbf {\bibinfo {volume} {101}},\ \bibinfo {pages} {052401} (\bibinfo {year} {2020})}\BibitemShut {NoStop}%
\end{thebibliography}%

\end{document}